\documentclass[11pt,onecolumn,a4paper]{article}
\usepackage[latin1]{inputenc}
\usepackage{epsfig}
\usepackage[english]{babel}
\usepackage{textcomp}
\usepackage{amsmath}
\usepackage{amsfonts}
\usepackage{amssymb}
\usepackage{float}
\usepackage{xcolor}
\usepackage{abstract}
\usepackage{array,multirow}
\usepackage{graphicx}
\usepackage{subcaption}
\usepackage{authblk}
\usepackage{color, colortbl}

\title{\bfseries Charge coupling in multi-stage laser wakefield acceleration}
\author{$\rm N. ~Pathak^{1,2}$\thanks{naveenpathak@sanken.osaka-u.ac.jp}}
\author{$\rm A. ~Zhidkov^{1,2}$}
\author{$\rm Y. Sakai^{1,2}$}
\author{$\rm Z. Jin^{1,2}$}
\author{$\rm T. ~Hosokai^{1,2}$}

\affil{$\rm ^{1}$ \small {\em Institute of Scientific and Industrial Research (ISIR), Osaka University, Mihogaoka 8-1, Ibaraki, Osaka, 567-0047, Japan.}}
\affil{$\rm ^{2}$\small {\em Laser Accelerator R$\&$D , Innovative Light Sources Division, RIKEN SPring-8 Center, 1-1-1, Kouto, Sayo-cho, Sayo-gun, Hyogo, 679-5148, Japan}}
\date{}
\begin{document} 
\maketitle

\begin{abstract}
The multi-stage technique for laser driven acceleration of electrons become a critical part of full-optical, jitter-free accelerators. Use of several independent laser drivers and shorter length plasma targets allows the stable and reproducible acceleration of electron bunches (or beam) in the GeV energies with lower energy spreads. At the same time the charge coupling, necessary for efficient acceleration in the consecutive acceleration stage(s), depends collectively on the parameters of the injected electron beam, the booster stage, and the non-linear transverse dynamics of the electron beam in the laser pulse wake. An unmatched electron beam injected in the booster stage(s), and its non-linear transverse evolution may result in perturbation and even reduction of the field strength in the acceleration phase of the wakefield. Analysis and characterization of charge coupling in multi-stage laser wakefield acceleration (LWFA) become ultimately important. Here, we investigate acceleration of an externally injected electron beam in laser wakefield, emulating a two-stage LWFA, via fully relativistic multi-dimensional particle-in-cell simulations, and underlying the critical parameters, which affect the efficient coupling and acceleration of the injected electron beam in the booster stage. 
\end{abstract}

\newpage

\section*{Introduction}
Developing of laser driven plasma accelerators has shown significant progress in the last two decades \cite{Joshi,Esarey,VMalka,SMHooker} and now full-optical accelerators are considered as a new, promising lineup of the particle accelerators. Small sizes and low maintenance cost, affordability to small scientific laboratories make such accelerators attractive. 

In the frame of laser wakefield acceleration, electron energies over 8 GeV in a single stage has been demonstrated \cite{Gonsalves}. With the continuous progress in the high power laser technology, 10 GeV electron energies will be achieved in a single stage LWFA in near future \cite{WPLeem}. Fundamentally, in a single stage LWFA the maximum energy gain by an electron beam is limited by (i) pump depletion length, where the laser energy is no longer sufficient to drive a strong wakefield, and (ii) dephasing  length, where the electron beam ultimately runs out of phase with the accelerating plasma wave \cite{Esarey}. Besides, maintaining the stable propagation of the laser pulses over long (several tens of centimeters or meter scale) plasma is a very challenging issue, due to various instabilities \cite{Kruer,Tripathi,Gibbon} and technical problems in providing very long uniform guiding structures. 

Multi-stage LWFA schemes provide a solution for avoiding the maximum energy gain limitation in a single stage \cite{LeemanEsarey}. It may also help in reducing the instabilities that may gradually arise in a long distance propagation of the laser pulses in a single stage. The electron beam can be accelerated successively in many stages, and eventually very high-energy gain could be achieved \cite{Luo,Xiangkun}. The technical requirements in multi-stage LWFA are: (i) synchronization (or coupling) between the successive acceleration stages, and (ii) efficient transportation of the electron beams from one stage to another stage with high transparency. These necessities, which are also a critical part in vacuum accelerators, are extensively studied for LWFA \cite{Panasenko,Tilborg,Sakai} and successfully implemented in some preliminary staging LWFA experiments \cite{Steinke,Jin}. 

The potential key issues that are important for the performance of multi-stage LWFA is its efficiency \cite{comment} in providing the high charge coupling and energy gain to the injected electron beam along with necessary low energy spread. These conditions require that the booster stage parameters must match well with the parameters of the electron beams that are generated from the LWFA based cathode: injector stage, and vice versa. For example, in order to achieve hundred percent charges coupling efficiency the transverse size of the injected electron beam must be less than the transverse size of the wakefield. And in order to accelerate the entire electron beam its longitudinal size should match with the longitudinal size of the accelerating phase of the wakefield ($\lambda_{p}/2$), where$\lambda_{p}$ is the plasma wavelength. The transverse size of the wakefield is of the order of the laser spot size ($2w_{0}$), whereas, the longitudinal size of the acceleration field is determined by the plasma electron density: smaller the plasma electron density ($N_{e}$), longer the accelerating wake wave. At high intensity the plasma wavelength also depends on laser intensity and is longer than the linear plasma wavelength \cite{Esarey}. 

The maximum energy gain by an electron beam is limited by dephasing length ($L_{d}$), which is inversely proportional to the plasma density as: $L_{d} \propto \frac{1}{N_{e}^{3/2}}$ \cite{Esarey}. Therefore, in order to achieve a higher energy gain of the electron beam, the plasma density should be reduced. Similarly, the critical power ($P_{cr}$) required for self-guiding of the laser pulse is also inversely proportional to the plasma density: $P_{cr} (W)= 1.7\times 10^{10}\frac{N_{cr}}{N_{e}}$ \cite{Esarey,Gibbon}. Therefore, in the case of uniform density plasma and moderate laser powers, self-guiding (or self-focusing) of the laser pulse, allowing the longer propagation length in plasma, requires high electron density $N_{e} > \frac{1.7\times10^{10}N_{cr}(cm^{-3})}{P(W)}$, where $N_{cr}$ is the critical density for the laser pulse frequency and $P$ is the total power of the laser pulse. Thus, the minimum plasma density requirement for self-guiding of the laser pulse in uniform density plasma determines a limit on the longitudinal size ($\lambda_{p}\propto \frac{1}{\sqrt{N_{e}}}$) of the accelerating field for a given laser power.

The transverse and longitudinal sizes of the electron beam are determined by beam's geometrical emittance, mean energy and energy spread $\triangle \gamma$ ($\gamma$ is the relativistic factor: $\gamma=\sqrt{1+(p/mc)^{2}}$. If the distance between an injector and a booster stage is $L$, the bunch length at the entrance point of the booster stage will be equal to $l=L\triangle\gamma / \gamma_{0}^{3}$ where $\gamma_{0}$ is the mean energy of electrons in the bunch (We assume $\triangle \gamma \ll \gamma_{0}$ and $l$ is bigger than the initial bunch length). In case of $l > (\lambda_{p}/2)$ an essential portion of the electron beam cannot be injected in the acceleration phase of the laser wakefield. Thus, in order to match the electron beam length with the accelerating wakefield the beam should have low energy spread, and must satisfy the minimum mean energy requirement condition. Moreover, the longitudinal and transverse space charge effect scale as $1/(\beta^{2}\gamma^{3})$ and $1/(\beta^{2}\gamma^{5})$, respectively \cite{Reiser}. Therefore, by increasing the electron energy larger number of electrons can be compressed in a short bunch. However, the influence of high bunch charge density on the longitudinal component of the wake may further limit the acceleration process in the booster stage, which inturn spoil the beam quality, in particular, energy spread. 

In ref \cite{Reitsma} post-acceleration of an electron bunch with an exponential energy distribution, $f(\gamma) \propto \exp(-\gamma m_{e}c^{2}/T_{e})$, was studied numerically,  where $T_{e}$ is an effective electron temperature in the MeV range. In ref \cite{stageMalka} energy gain of an externally injected relativistic as well as sub-relativistic electron beam in plasma channel was demostrated numerically. However, due to the bunch length elongation criteria ($l=L\triangle\gamma / \gamma_{0}^{3}$) neither large energy spread nor sub-relativistic electron beams are suitable for practical staging experiment. Similarly, modeling of an externally injected electron beam inside a gas filled dielectric capillary tube is presented in \cite{Bhusan}. Although all these studies are important numerical initiatives towards the multi-stage LWFA configuration, none of these quantify how to determine the parameters of an injected electron beam and the booster stage. % \textcolor{red}{in order to achieve high energy gain for a dense electron beam keeping the low energy spread in multi-stage LWFA configuration}.

It is well known that the space-charge effect of an externally injected electron beam will influence the longitudinal electric field strength of the wake. To understand when the electron beam space-charge provoke nonlinear deterioration to the longitudinal electric field of the wake, let us assume a reference frame moving with the initial velocity of the injected electron beam. In this frame the amount of the bunch charge that can be accelerated efficiently in the booster stage can be roughly estimated by comparing the longitudinal component of the electric field strength of the self-induced electric field of the beam with the longitudinal component of the field strength of the wake. In this reference frame the longitudinal component of the self-induced electric field of the injected beam is given by $\partial E_{x}/\partial x = -4\pi en_{b}$, where $n_{b}$ is the electron beam density. For illustration, we can write $E_{x} = -4\pi en_{b}L$, where $L$ is the electron beam length. The electron beam density can be given as $n_{b}=\frac{N_{eb}}{(2\pi )^{3/2}r_{b}^{2}L}$, where $N_{eb}$ is the total number of electrons in the bunch, $r_{b}$ is the electron beam radius. Therefore, $E_{x}=-4\pi e\frac{N_{eb}}{(2\pi )^{3/2}r_{b}^{2}}$, or $E_{x}\sim\frac{Q}{r_{b}^{2}}$, where, $eN_{eb}=Q$ is the electron beam charge. The longitudinal component of the wakefield remains unchanged in this frame. The longitudinal electric field strength of the plasma wake excited by the laser pulse, ($E_{p}$), is proportional to the laser electric field ($E_{L}$) and plasma density ($N_{e}$) as: $E_{p} \sim E_{L}\frac{\omega_{p}}{\omega_{L}}$ \cite{Bulanov,YGu}, $\omega_{p}$ is the plasma frequency and $\omega_{L}$ is the laser frequency. In normalized unit the electric field strength of the laser pulse can be written as $E_{L}=\frac{a_{0}m_{e}c\omega_{L}}{e}$ , where $c$ is the speed of light and $a_{0}$ is the normalized vector potential. Therefore, the electric field strength of the plasma wave excited by the laser pulse can be written as $E_{p}\sim \frac{a_{0}m_{e}c\omega_{p}}{e}$. In order to avoid any strong perturbation of the longitudinal component of the electron beam field strength on the longitudinal field strength of the wake excited by the laser pulse $\frac{Q}{r_{b}^{2}} \ll \frac{a_{0}m_{e}c\omega_{p}}{e}$, or $Q \ll (a_{0}\omega_{p}r_{b}^{2})$. This ad-hoc estimation implies that for an efficient acceleration of the injected electron bunch in the booster stage the amount of bunch charge is limited by: (i) laser pulse intensity, (ii) plasma density and (iii) electron beam radius. Moreover, the electron beam radius is a dynamically evolving parameter due to the transverse focusing force of the plasma wave, and partially due to the self-generated azimuthal magnetic field. Therefore, the coupling not only includes the electron beam charge, but also includes the electron beam density that may not remain constant during the acceleration process. 

The purpose of this paper is to investigate the coupling and acceleration efficiency of an externally injected electron beam with the booster stage plasma waves in real geometry, via multi-dimensional fully relativistic particle-in-cell simulations. Since the coupling and evolution of an electron beam (or beam density) in the booster stage is a dynamic process, the ad-hoc estimation is only qualitative. Proper understanding of the coupling process require an ab-initio simulations. This allows us to study the full nonlinear evolution of the electron beam and wakefield depending on the laser, plasma and injected electron beam parameters. And pave path to underlying the critical parameters, which affect the coupling and acceleration of an electron beam in the booster stage in multi-stage LWFA scheme. Problem of emittance correction of an externally injected electron beam with the beam optics \cite{Jin} is not considered in this work.

\section*{Simulation results}

The numerical study of coupling of an externally injected electron beam with the LWFA based booster stage is performed both in 3D and 2D geometry using relativistic particle-in-cell (PIC) code FPLaser \cite{Zhidkov1,Zhidkov2}. Except Fig. 9 and Fig. 10 all simulations were performed in 3D geometry. The simulations were performed using moving window technique. In the simulations the laser pulse propagates along $x$-direction [from the right hand side (RHS) to the left hand side (LHS)].  The parameters of the laser pulse, plasma density and injected electron beam are chosen considering the requirements for a single-stage (or booster-stage) LWFA with an external injection: emulating the two stage acceleration. Foremost, in order to have large dephasing length we consider a low-density plasma medium, $5\times 10^{17}cm^{-3}$, for the booster stage. The fully ionized underdense plasma has a linearly increasing density profile along the propagation direction up to $50 \ \mu m$ and then constant. The energy boosting of the electron beam in uniform density plasma as well as parabolic plasma channel is studied. In the case of parabolic plasma channel the profile of the plasma density is $N_{e}(r)=N_{emin}+\triangle N_{e}(r^{2}/R^{2})$, where $R$ is the radius of the channel, $N_{emin}$ is the on axis plasma density, and $\triangle N_{e}$ is the density depth. The size of the simulation window is $x \times y \times z=(120 \times 300 \times 300)\mu m^{3}$ in 3D geometry, and $x \times y=(120 \times 300)\mu m^{2}$ in 2D geometry. The moving simulation box was kept long enough in order to simulate at least two full plasma buckets behind the laser pulse and to capture (if any) the dynamics of the effect of the injected electron beam. In 3D geometry the axial resolution is $\lambda_{0}/64$ and the transverse resolution is $\lambda_{0}/16$ and $8$ particles per cell, whereas, in 2D geometry the axial resolution is $\lambda_{0}/64$ and the transverse resolution is $\lambda_{0}/32$ and $25$ particles per cell. High spatial and temporal resolution has been chosen to diminish the effects of numerical dispersion. The driver laser pulse parameters for the booster stage are chosen for a standard Ti:Sapphire based laser system. We consider a laser pulse of central wavelength $\lambda_{0}=800 \ nm$ and full width half maximum (FWHM) pulse duration of $\tau = 30 \ fs$. The laser pulse energy is $\rm E_{L}= 4 J (133 \ TW)$ in order to have pulse power at least two time greater than the critical power for self-focusing. The laser spot size is chosen to satisfy the matching condition, $k_{p}w_{0} \simeq 2\sqrt{a_{0}}$, that implies $2w_{0} = 40 \ \mu m$ (diameter at the $1/e^{2}$ intensity position), which corresponds to the vacuum intensity of $I=1\times 10^{19}\ Wcm^{-2} \ (a_{0}=2.16)$. The best focus position is located at the entrance of the simulation box. The parameters of an externally injected electron beam are chosen based on the matching criteria with the accelerating field in the booster stage (as discussed in the introduction part). Considering a typical distance of around one meter from the injector stage to the booster stage, the mean energy of the electron beam is chosen to be $E_{B}=100 \ MeV$ with an energy spread of $ \frac{\triangle E_{B}}{E_{B}}=3\%$ . The spatial and temporal profile of an externally injected electron beam is Gaussian with diameter $10 \mu m$ and length $10\mu m$. The delay between the booster laser pulse and electron beam is $100 \ fs$ in order to overlap the electron beam with the accelerating phase of the wakefield in the first plasma bucket. To estimate the charge coupling and acceleration efficiency, a wide range of electron beam charge $(10-100) \ pC$ is used in the simulations. 

The prime goal was to investigate: (i) how much electron beam charge can be efficiently coupled with the booster stage within a given booster stage parameters, (ii) implications of the charge coupling on booster stage as well as on the electron beam parameters, and (iii) how the electron beam charge coupling efficiency with the booster stage can be increased. 

Beginning of the process of coupling and acceleration of an externally injected electron beam in the booster stage is illustrated by Fig.1 presenting results of 3D PIC simulations. Fig.1 (a) shows the laser pulse propagating in uniform underdense plasma in the booster stage followed by an externally injected electron beam. Fig. 1(b), and 1(c) show the evolution of plasma wake and trapping of the electron beam in the first plasma bucket behind the laser pulse. Fig. 1(d) shows 3D  illustration of the trapping and acceleration of an externally injected electron beam in the booster stage.

Fig.2 shows coupling effect of an electron beam charge on the wakefield of the booster stage. Fig. 2(a), 2(b), 2(c), 2(d) and 2(e) corresponds to $10 \ pC$, $20 \ pC$, $30 \ pC$, $50 \ pC$ and $100 \ pC$, respectively. These results are shown at 2.5 ps simulation time or after $750 \rm \mu m$ propagation length. In this time scale self-consistent evolution of the electron beam (periodic focusing and defocusing) can be clearly seen in the simulations. The dynamic evolution of the electron beam is shown in the next Section. In Fig.2, red-white-blue colormap shows the normalized axial or longitudinal filed of the plasma wave. The blue region is the decelerating part of the plasma wave, whereas, the red region is the accelerating part of the plasma wave. The multi-color map shows the electron beam density normalized by critical plasma density. The 1-D line out in the figure display the normalized effective on-axis wake electric field, which is the result of superposition of the longitudinal electric field of electron beam and plasma wave. With the increasing electron beam charge, the effective electric field of the wake decreases or deteriorated non-linearly. Moreover, the scale length of the deterioration of the effective accelerating field is of the order of electron beam length. This will spoil the electron beam characteristics. As noticed from Fig.2, for the parameters in the simulations, this effect is more prominent for beam charge $\geq 30 \ pC$. 

Fig. 3 shows the effect of this non-uniform electric field of the wake on the energy spectrum of the accelerated electron beam in the booster stage. As observed from Fig. 2(b-c), for $20$ and $30 \ pC$ bunch charge the amplitude of the net wake electric field is decreased and modulated in a manner that the front part of the electron bunch is accelerated by higher field, whereas, the rear part is accelerated by relatively lower field. As a result of which the energy spectrum of the electron beam starts broadening [Fig. 3(b-c)]. In Fig. 2(d), for $50 \ pC$ bunch charge the net field at the rear part of the electron bunch become decelerating. Consequently, the rear part of the electron beam experiences decelerating force and lose energy [Fig. 3(d)]. This is a serious problem with high bunch charge. It may completely spoil the purpose of multi-stage acceleration scheme. For higher bunch charge of $100 \ pC$ this is even more serious, where a significant portion of the electron bunch experiences decelerating force [Fig. 2(e)] and deceleration of beam particles dominates over acceleration [Fig.3(e)]. Even though similar beam loading effects was first reported experimentally in single stage LWFA \cite{Rechatin}, its characterization in multi-stage LWFA scheme is quite important where electron beams are expected to gain higher energies in the few GeV range keeping low energy spread and high charge.  

Deterioration of the accelerating plasma field with increasing bunch charge is a serious limitation in LWFA, in particular in the linear regime, and implies that acceleration of dense electron beams with unprecedented qualities require properly optimize multi-stage schemes. 

\section*{Dependence of charge coupling efficiency on laser pulse intensity, plasma density and electron beam size}

In this Section we investigate the dependency of acceleration efficiency of $50 \ pC$ charge on laser pulse intensity and plasma density in the booster stage, and transverse size of the injected electron beam. The laser pulse intensity could be increased either by reducing the focus spot, pulse duration or by increasing the pulse energy. Small laser spot require smaller diameter of the injected electron beam, which increases the beam density. Pulse duration smaller than $30 \ fs$, and with few Joule energy require specially designed laser systems. The standard laser systems can deliver multi-Joule $30 \ fs$ laser pulses. Therefore, we choose to increase the laser pulse energy in order to gain higher intensity. The laser pulse energy is increased from $4 \ J$ to $10 \ J$, which corresponds to the pulse intensity of $I=2.5\times 10^{19} \ Wcm^{-2} \ (a_{0}=3.42)$. Except laser pulse energy or focused intensity all other parameters remain unchanged. Fig. 4 shows dependency of coupling and acceleration of $50 \ pC$ electron beam on laser pulse intensity in the booster stage. Like Fig. 2, the red-white-blue colormap shows the normalized longitudinal filed of the plasma wave, and the multi-color map shows the electron beam density normalized by critical plasma density. The comparison is shown at 2.5 ps. In Fig. 4(a) the 1D line plot in blue color shows an on-axis effective accelerating field for 10 J laser pulse, whereas, the 1D line plot in magenta color shows an effective accelerating field for 4 J laser pulse. In Fig. 4(b) blue color display the energy spectrum of an injected electron beam with 10 J laser pulse in the booster stage, whereas, magenta color display energy spectrum with 4 J laser pulse. These figures imply that the affect of beam-field of $50 \ pC$ electron bunch on the accelerating field can be suppressed by increasing the laser pulse intensity in the booster stage. This further indicates that for a fixed plasma density and electron beam size the maximum injected charge in the booster stage should scale with the intensity of the driving laser pulse in the booster stage.    

In the next case we check the dependency of the charge coupling on plasma density in the booster stage. Therefore, the plasma density in the booster stage is increased from $5 \times 10^{17}cm^{-3}$ to $1 \times 10^{18}cm^{-3}$. All other parameters remain unchanged. Fig. 5 shows comparison of acceleration of $50 \ pC$ electron beam for different plasma density in the booster stage. The comparison is shown at 2.5 ps. In Fig. 5(a) the 1D line plot in blue color shows an on-axis effective accelerating field for $1 \times 10^{18}cm^{-3}$, whereas, the 1D line plot in magenta color shows an effective accelerating field for $5 \times 10^{17}cm^{-3}$. In Fig. 5(b) blue color display the energy spectrum an externally injected electron beam with plasma density of $1 \times 10^{18}cm^{-3}$ in the booster stage, whereas, magenta color display energy spectrum with plasma density of $5 \times 10^{17}cm^{-3}$. The role of high plasma density in canceling out the influence of self-field of the electron beam is clearly visible. Due to higher plasma density, plasma wavelength is smaller. Therefore, the electron beam delay with respect to the laser pulse needs to be adjusted in order to achieve phase matching with the accelerating field. Moreover, in higher plasma density the self-focusing effect may also increase the local intensity of the laser pulse. As a result, the transverse focusing force of the plasma wave and dynamic evolution of the electron beam are different than those in the case of lower plasma density. The adverse effect of higher plasma density is a lower dephasing length, which may limit the maximum energy gain. 

Improvement in the acceleration of the dense electron beam in the booster stage, independently with higher intensity and plasma density, implies that it could be further enhance by integrating both these effects. Hence, we increase simultaneously the laser pulse energy as well as the plasma density in the booster stage. Now, the laser pulse energy is $10 \ J$ instead of $4 \ J$, and the plasma density is $1\times 10^{18} cm^{-3}$ instead of $5\times 10^{17} cm^{-3}$. The initial transverse size of the electron beam remain unchanged. Moreover, the lower dephasing length due to higher plasma density could be partially compensated by intensity dependent nonlinear increase in the plasma wavelength. Fig. 6(a) shows comparative evolution of the net accelerating field in this case. The comparison is shown at 2.5 ps. Blue color 1D line out shows net accelerating field for high intensity as well as plasma density, whereas, the magneta color 1D line out shows net accelerating field for relatively low intensity and plasma density. Fig 6(b) shows evolution of the electron beam energy in the booster stage. The blue color 1D line out shows the energy evolution for high intensity and plasma density. One can observe drastic improvement and enhancement of the net accelerating force, which results in rapid acceleration of the $50 \ pC$ electron beam as compared to the previous cases. Moreover, nearly all the electrons in the beam experiences accelerating force. This may partially help in preserving the electron beam qualities. Thus, the acceleration efficiency of $50 \ pC$ electron bunch enhanced significantly by properly choosing the laser pulse intensity and plasma density in the booster stage. 

Next, we examine the dependency of acceleration process on initial transverse size of the injected electron beam. In this case the initial electron beam radius ($r_{b}$) is increased from $5 \ \mu m$ to $10 \ \mu m$. All other parameters remain unchanged (laser pulse energy $4 \ J$ and plasma density $5 \times 10^{17}cm^{-3}$). Fig. 7 shows dependency of the acceleration of $50 \ pC$ electron beam on its transverse size. In Fig. 7(a) the 1D line plot in blue color shows an on-axis effective accelerating field for $r_{b}=10 \ \mu m$, whereas, the 1D line plot in magenta color shows an effective accelerating field for $r_{b}=5 \ \mu m$. In Fig. 7(b) blue color display the energy spectrum of an externally injected electron beam of radius $r_{b}=10 \ \mu m$, whereas, magenta color display energy spectrum of an externally injected electron beam of radius $r_{b}=5 \ \mu m$. This clearly implies that it is not the initial bunch charge but the bunch charge density that should match with the booster stage parameters. However, increasing the initial diameter of the electron beam partially require a larger transverse size of the wake field (or laser pulse focus spot in order to avoid beam loss), which, in turn, will require larger focal length of the laser pulse focusing optics. 

An important feature of the injected electron beam that needs to be address is its transverse dynamic evolution during the acceleration process in the booster stage. The electron beam is under the influence of focusing force of the accelerating plasma wave. Due to the focusing force the transverse size, and hence, the density of the electron beam vary periodically in the booster stage. Consequently, the corresponding net accelerating force also vary dynamically, which means the rate of acceleration doesn't remain constant during the acceleration process. Fig. 8 shows evolution of $r_{b} = 10 \mu m$ electron beam injected in the booster stage. Fig. 8(a-e) illustrate the dynamic focusing and defocusing of the electron beam at 0.4 ps, 0.6 ps, 0.8 ps, 1.0 ps and 1.5 ps, respectively. Fig. 8(f) shows the corresponding net accelerating field. For an ideal booster stage configuration, constant accelerating filed over the entire electron bunch length is required to avoid any further energy spread in the electron beam. In Fig. 8(f) the accelerating gradient is flat (or constant) at 0.4 ps and then vary periodically. In general, periodic evolution of an externally injected electron beam is indispensable in the plasma wave. The energy spread, if introduced in the booster stage, can be either partially compensated by simply dephasing the electron beam in the booster stage or by employing the plasma dechirper concept \cite{YWu}. 

Table 1 summarize the acceleration efficiency of $50 \ pC$ electron beam in the booster stage with different laser pulse intensity, plasma density and electron beam diameter. One can clearly see that the simulation results are in good agreement with the ad-hoc estimation: $Q \ll (a_{0}\omega_{p}r_{b}^{2})$, and suggest that for coupling and acceleration of a dense electron beam, generated either from LWFA based injector or from vacuum accelerator, require proper matching with the plasma and driving laser pulse parameters in the booster stage(s) in order to avoid any degradation in the beam properties. 

\section*{Refraction losses of laser pulse in uniform underdense plasma}

Another important physical aspect for energy gain of an electron beam from the plasma waves is stable excitation of wakefield, which in turn require stable propagation of the driving laser pulse in the booster stage. For a sufficiently high power laser the longitudinal ponderomotive force creates a density gradient in front of the leading edge of the pulse. This density gradient results in refractive index variation in front of the leading part of the laser pulse. Consequently, in the absence of any guiding medium the laser pulse gradually refract in the transverse direction. Due to refractive loses the laser pulse intensity gradually decreases, and hence the amplitude of the excited plasma wakefield. Fig. 9 (a) shows accelerating plasma field evolution in the booster stage with uniform density plasma. The laser pulse, plasma density and electron beam parameters are the same as for the case in Fig. 2(a). The critical power for self-focusing is $\sim 60 \ TW$. The power in the laser pulse is $133 \ TW$. In-spite of the fact that the laser power is greater than the critical power for self-focusing, it cannot maintain the self-focused propagation for longer length due to density gradient induced refraction of the laser pulse. This imposes limitation on the efficient acceleration and maximum energy gain of the electron beam in the booster stage. To suppress the refractive loses proper guiding of the laser pulse is required. Therefore, in the booster stage guiding of the laser pulse in pre-formed plasma channel created either by optical methods \cite{Lemos,Shalloo} or by capillary discharge \cite{Hosokai,Butler} is desirable. Fig. 9(b) shows accelerating plasma field evolution in pre-formed plasma channel. All parameters in the simulation are same as in the case of uniform density plasma except, now plasma channel is used for guiding of the laser pulse. The plasma channel has a parabolic profile in the radial direction. We choose a shallow plasma channel of an on axis plasma density of $5\times 10^{17}cm^{-3}$, density depth of $\triangle N_{e}=0.5$ and diameter $100 \ \mu m$. Fig. 9(b) shows that even a shallow plasma channel is quite effective in preventing the refractive loses, and may ensure efficient coupling as well as energy gain of a well-matched electron beam in the booster stage. Fig. 10(a) shows the energy spectrum of the electron beam co-propagating with the plasma wave in the booster stage, which consist of uniform density plasma. Whereas, Fig. 10(b) shows energy spectrum in the booster stage consist of parabolic plasma channel. The energy spectrum in both cases is shown at $30 \ ps$ or after $9 \ mm$ propagation length in the booster stage. Acceleration gradient reduction and low energy gain due to refraction loses is apparent from Fig 9(a). Moreover, due to low plasma density indeed the dephasing length is longer, but the energy gain per millimeter propagation length is not high enough even in plasma channel. Very long propagation length is required in order to gain quite high energies. Therefore, it is important to optimize all the parameters in the booster stage, which not only allow high charge coupling but also enable high-energy gain per 'boosting-stage' in multi-stage LWFA configuration. 

\section*{Design requirements for stable, high charge and high energy multi-stage LWFA}
Necessity for multi-stage LWFA configuration basically comes from the high-energy gain limitation in a single stage LWFA, as well as the stability and reproducibility problems in a very long single stage plasma (or plasma channel). For real practical applications in addition to high energy, high charge and low energy spread is also necessary. For instance, high brightness electron beams with narrow energy spread are essential for driving short wavelength, high gain Free Electron Laser (FEL) \cite{Simone}, or high luminosity electron beams are essential for future LWFA based collider schemes \cite{Schroeder,Nakajima}. Therefore, design approach for multi-stage LWFA configuration should include high charge, high-energy gain as well as narrow energy spread.

For high brightness and low emittance electron beams, its transverse size must be small enough. Thus, the design for multi-stage LWFA is mainly focused on optimizing the intensity and plasma density in the booster stage(s). The advantage of low-density plasma is longer dephasing length. Similarly, advantage of low intensity laser pulse driver is near instability free or quasi-stable propagation (including plasma channel). However, dis-advantages of low intensity laser pulse and low-density plasma in the booster stage are (i) low energy gain per unit acceleration length, and (ii) acceleration of low charge. As the energy gain is low, it requires very large acceleration length (long plasma) to reach high energy. Stability and reproducibility of such a long plasma (or plasma channel) is a crucial technical issue.

On the other hand, at high intensity and high plasma density energy gain of the injected electron beam per unit acceleration length is higher. As a result, higher energy gain is achieved in relatively smaller acceleration length. Further, the charge coupling efficiency is also high. In longer propagation length, high intensity and high plasma density may cause various instabilities. A possible solution to this problem is sacrificing a part of the total useful acceleration length (to avoid instabilities) and increasing the number of booster stages. It is noteworthy that even after sacrificing a part of useful acceleration length, due to higher energy gain per unit length, the final energy is comparative to longer acceleration length in low density plasma. This indeed will require more laser pulses (or splitting of a powerful laser pulse into multiple beam lines), but will ensure the stability and reproducibility of "high brightness" and "high energetic" electron beams with unprecedented beam qualities. 

\section*{Discussion}
Demonstration of stable and reproducible acceleration of high charges ($\sim$ hundreds of pC) to high energies ($\sim$ multi-GeV class) with low energy spread ($\leqslant 1\%$) is one of the main goal in LWFA scheme. Considering the low energy spread requirement, linear acceleration regime ($a_{0} \sim 1$), which is less prone to instabilities is limited to acceleration of lower charges (few tens of $\sim$ pC) \cite{Katsouleas}, whereas, nonlinear regime, which allows acceleration of higher charges is prone to various instabilities in longer acceleration length. Therefore, proper optimization of the entire acceleration process is essential to deliver high quality beams that can be used for potential scientific and industrial applications. Nonlinear multi-stage LWFA scheme could provide solution to this problem.

In the case of self-injection scheme the wakefield evolve consistently to its wave breaking limit and the loaded charge is gradually increases over hundred's of femto-seconds time scale. Whereas, in the case of external injection scheme the amount of charge is fixed and influence the wakefield in the booster stage from the very beginning of the interaction process. Therefore, the nonlinear process may not occur similarly in both the cases. It is noteworthy, that the amount of charges accelerated efficiently in the simulations presented here do not agree well with the existing nonlinear scaling for self-injection regimes. 

In the non-linear case, using matching conditions the number of particles that can be loaded into a three-dimensional wake by self-injection process was estimated by Lu et.al. \cite{WuLu}, $N\simeq \frac{8/15}{k_{0}r{e}}\sqrt{\frac{P}{mc^{3}/r_{e}}} \simeq 2.5\times 10^{9}\frac{\lambda_{0}(\mu m)}{0.8}\sqrt{\frac{P(TW)}{100}}$. The similar scaling was also obtained by Gordienko et.al. \cite{Gordienko}, $N\simeq \frac{1.8}{k_{0}r{e}}\sqrt{\frac{P}{mc^{3}/r_{e}}}$, where $r_{e}=e^{2}/(mc^{2})$ is the classical electron radius, $k_{0}=2 \pi /\lambda_{0}$ is the laser wave number, and $P$ is the power of the laser pulse. Optimal loading of the wakefield with tailored electron bunch in the bubble regime has been also investigated by Tzoufras et. al. \cite{Tzoufras}. This scaling also gives similar estimation as Lu et. al., the accelerated bunch charge is proportional to the square root of laser power $Q \propto \sqrt{P}$. In accordance with Lu et.al., in a non-linear regime for a $100 \ TW$, $0.8 \ \mu m$ laser pulse the number of accelerated electrons are $\sim 400 \ pC$. Thus, this scaling implies that in a self-injected single stage LWFA, for a matched laser pulse, few hundred $pC$ charge is expected to be accelerated efficiently for laser power $\geq 10 \ TW$. However, in contrast we observed quite less charge that could be accelerated efficiently in external injection case. The laser pulse power in the simulation presented in this manuscript is $133 \ TW$ and laser wavelength is $0.8 \ \mu m$. Moreover, the laser and plasma parameters in the booster stage satisfy the matching condition, $k_{p}w_{0} \simeq 2\sqrt{a_{0}}$. In-spite of these parameters and matching conditions, it was found that acceleration of even $30 \ pC$ charge is not efficient in the booster stage. The simulation results presented in this manuscript imply that for accelerating few hundred $pC$ to $nC$ level of electron bunch in the two-stage LWFA configuration, with unprecedented beam qualities (charge, energy spread, energy gain), high power laser pulses scaling from few hundred terawatt up to petawatt levels are required. However, perfect focus-ability of such high power laser pulses are hardly achievable, and therefore, halo problem \cite{Nakanii,Pathak} will be an another key issue for multi-stage LWFA schemes. 

\section*{Conclusions}
Multi-stage LWFA configuration is important for determining the future of laser driven plasma accelerators for achieving high energetic electron beams relevant to high energy physic. Proper understanding of coupling between each stage for efficient acceleration and energy gain of the electron beams is essential for its successful implementation. High energy, high charge and low energy spread is required for practical applications. We have presented comprehensive analysis of coupling and acceleration of an externally injected electron beams with LWFA based booster stage; emulating the two-stage LWFA scheme. The ab-initio fully relativistic PIC simulations delineate the influence of space-charge effect of an externally injected electron beam on the longitudinal electric field of the wake. Dependency of maximum charge coupling and efficient acceleration in the booster stage on laser pulse intensity, plasma density and electron beam size is investigated. The simulations results suggest that linear acceleration regime is limited to acceleration of few tens of pC charge preserving necessary low energy spread ($\leqslant 1\%$), whereas, acceleration of dense electron beams ($>$ 100 pC) to few GeV energy range and narrow energy spread ($\leqslant 1\%$) may need to operate in nonlinear acceleration regime. Properly designed multi-stage LWFA configuration may potentially overcome the limitations in single-stage LWFA: problem of long plasma channel uniformity, instabilities (filamentation, hosing etc.), limited energy gain, and acceleration of high charge at the cost of high energy spread. 

\section*{Acknowledgement} 
This work is funded by the JST-MIRAI program grant no. JPMJMI17A1, and was partially supported by the ImPACT R$\&$D Program of Council for Science, Technology and Innovation (Cabinet Office, Government of Japan). We are grateful to Prof. Yuji Sano, and Dr. Kando's group of QST-KPSI for their encouragements and helpful discussions. We also acknowledge the use of Mini-K computing facility at SACLA, RIKEN, SPring-8 Center. 

%===========================================================================
%\newpage

\newpage

%\begin{figure}[!tb]
%\centering
%\subfigure[]
%{\includegraphics[width=7.0cm]{channel_Ex_xy0525_6b}}
%\subfigure[]
%{\includegraphics[width=7.0cm]{channel_Ex_xy0525_6b}}
%\caption{ Reconstructed plasma density for the negatively chirped laser pulse using relativistically corrected Raman scattering relation. The laser pulse intensity is assumed to be constant. (a) the actual plasma density is $5 \times 10^{18} \ cm^{-3}$, and (b) the actual plasma density is $1 \times 10^{19} \ cm^{-3}$. Dashed line shows reconstructed plasma density obtained from the Raman shifted line and actual instantaneous laser pulse intensity in the simulations.}
%\label{fig.1}
%\end{figure}

\newpage

\begin{table}
%\centering
\begin{center}
\begin{tabular}{|c|c|c|c|c|c|c|c|}\hline
\multicolumn{8}{c}{Laser focus spot ($\rm 2w_{0}=40 \mu m$) and pulse duration ($\rm \tau=30\ fs$) are fixed} \\ \hline \hline
\multirow{2}{2.0cm}{Beam charge    (pC)} 
	& P & \centering $ \rm a_{0}$ & \centering $\rm N_{e}$ & $\rm r_{b}$ & $\rm E_{gain}$ & $\rm \triangle E$ & T \\
	& \centering ($\rm TW$) &  & \centering ($\rm cm^{-3}$) & \centering ($\rm \mu m$) & \centering (MeV) & \centering (FWHM, $\%$) & (ps) \\ \hline
\multirow{1}{*}{50} 
    & 133 & 2.16  &$\rm 5\times 10^{17}$ & 5 & 5 & 20.3 & 2.5 \\ \hline  
\multirow{1}{*}{50} 
    & 333 & 3.42 & $\rm 5\times 10^{17}$ & 5 & 20 & 7.3 & 2.5 \\ \hline       
    \multirow{1}{*}{50} 
    & 133 & 2.16 & $\rm 1\times 10^{18}$ & 5 & 15 & 8.2 & 2.5 \\ \hline
    \rowcolor{cyan}
    \multirow{1}{*}{50} 
    & 333 & 3.42 & $\rm 1\times 10^{18}$ & 5 & 40 & 3.5 & 2.5 \\ \hline
    \multirow{1}{*}{50} 
    & 133 & 2.16 & $\rm 5\times 10^{17}$ & 10 & 15 & 3.3 & 2.5 \\ \hline  
\end{tabular}
\end{center}
\caption{Summary of the acceleration efficiency of an externally injected $50 \ pC$ electron beam in the booster stage with different laser, plasma and electron beam parameters. Initial energy of the injected bunch is $100 \ MeV$ and energy spread $3 \%$. Here, $P$ is the laser power in terawatt, $a_{0}$ is the normalized laser pulse amplitude, $N_{e}$ is the plasma density, $r_{b}$ is the radius of the injected electron beam, $E_{gain}$ is the central energy gained by the electron beam in the booster stage, $\triangle E$ is the energy spread of the electron beam in the booster stage, and $T$ is the simulation time. The colored highlighted row shows the acceleration in the booster stage with high laser intensity and plasma density, which result in rapid energy gain with relatively low energy spread.}
\end{table}

\newpage

%=Figure:1===================================================

\begin{figure}
\centering
\includegraphics[width=\linewidth]{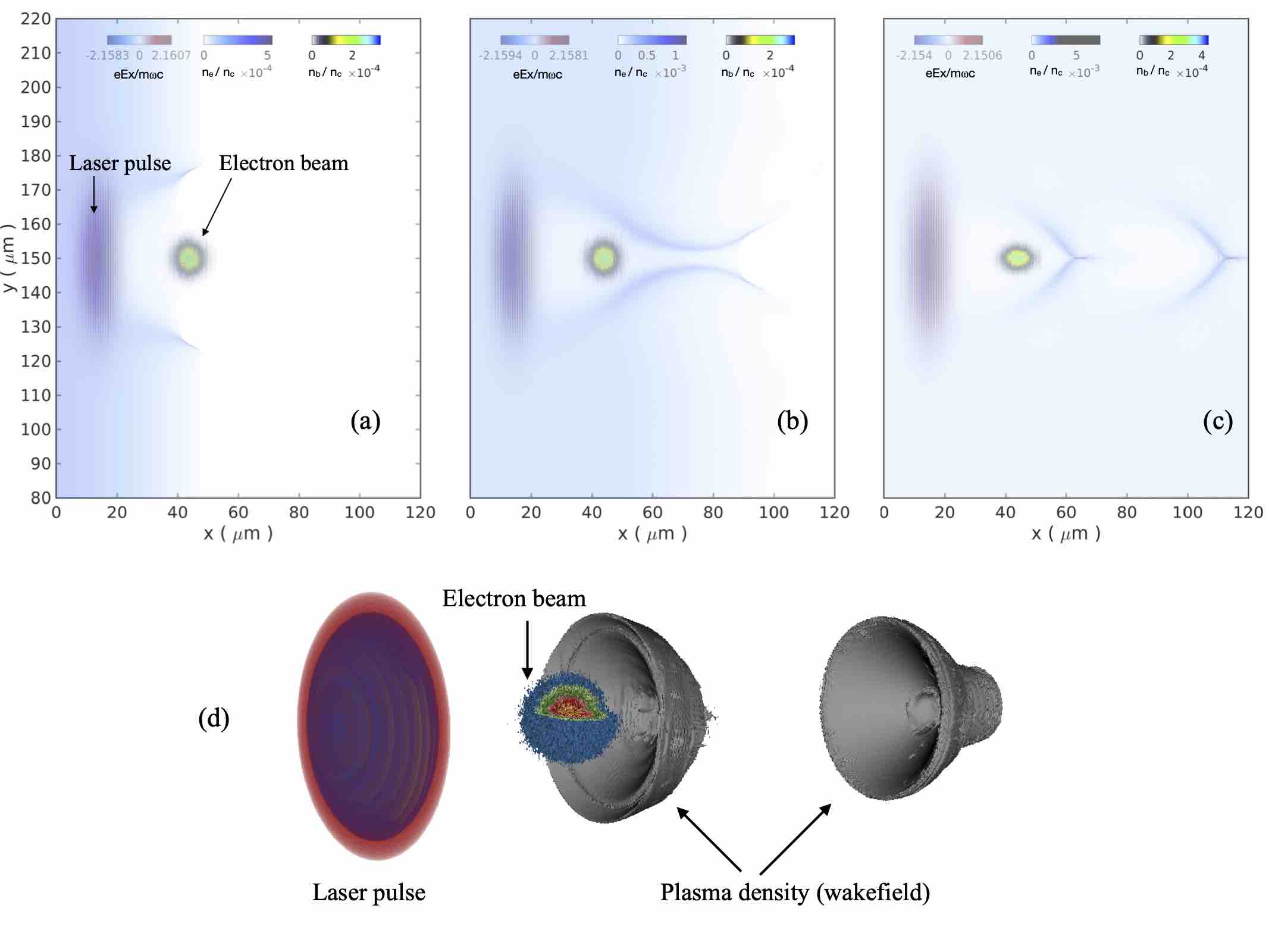}
\caption{\small Result of 3D PIC simulation illustrating the process of coupling and acceleration of an externally injected electron beam with the booster stage in two-stage LWFA configuration. (a) Shows the laser pulse is propagating in a uniform plasma density in the booster stage followed by an electron beam generated from an injector stage. (b) and (c) shows evolution of the plasma wave and trapping of the electron beam in the first plasma bucket behind the laser pulse. (d) 3D illustration of the PIC simulation showing the trapping and acceleration of an externally injected electron beam in the booster stage. The laser pulse is propagating from right to left hand side. The colorbar shows laser pulse amplitude in normalized unit ($eEy/m\omega c$), plasma density ($n_{e}$) normalized by critical density ($n_{c}$), and electron beam density ($n_{b}$) normalized by critical density.}\label{fig:1}
\end{figure}

%=Figure:2===================================================

\begin{figure}
\centering
\begin{subfigure}[t]{.4\textwidth}
\centering
\includegraphics[width=\linewidth]{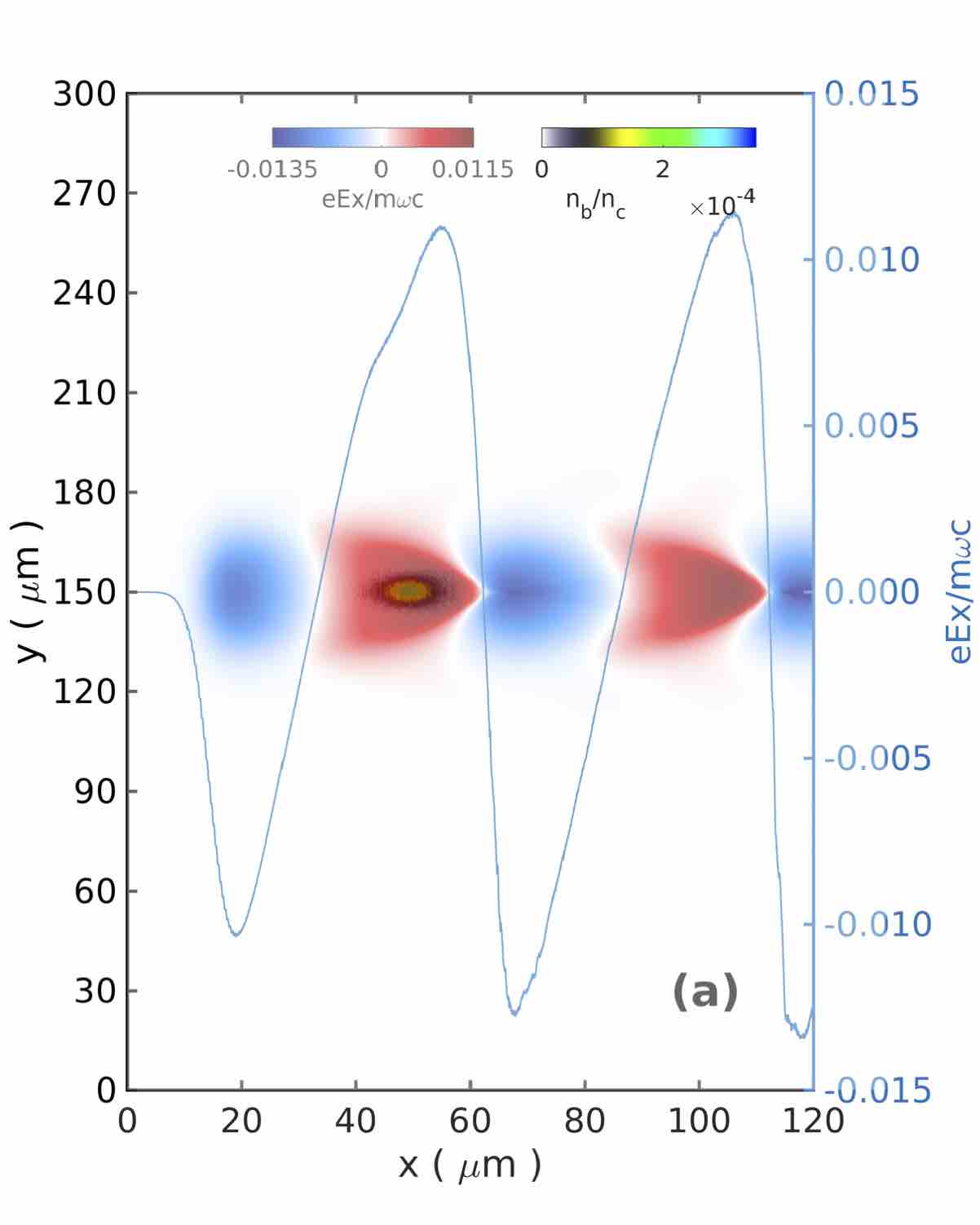}
\caption{}\label{fig:fig_a}
\end{subfigure}
\hskip 20pt
\begin{subfigure}[t]{.4\textwidth}
\centering
\includegraphics[width=\linewidth]{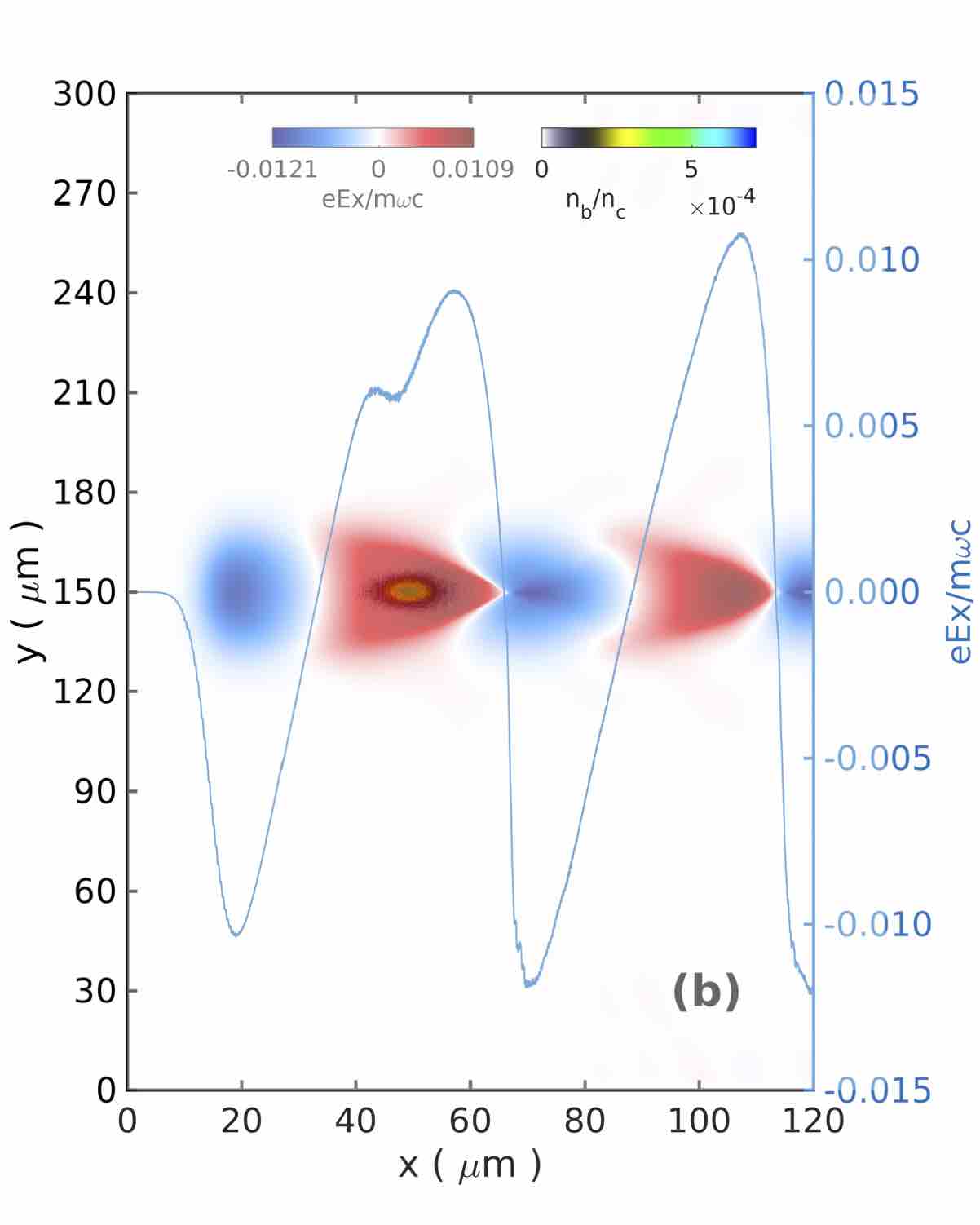}
\caption{}\label{fig:fig_b}
\end{subfigure}

%\medskip

\begin{subfigure}[t]{.4\textwidth}
\centering
\vspace{-20pt}% set the real top as the top
\includegraphics[width=\linewidth]{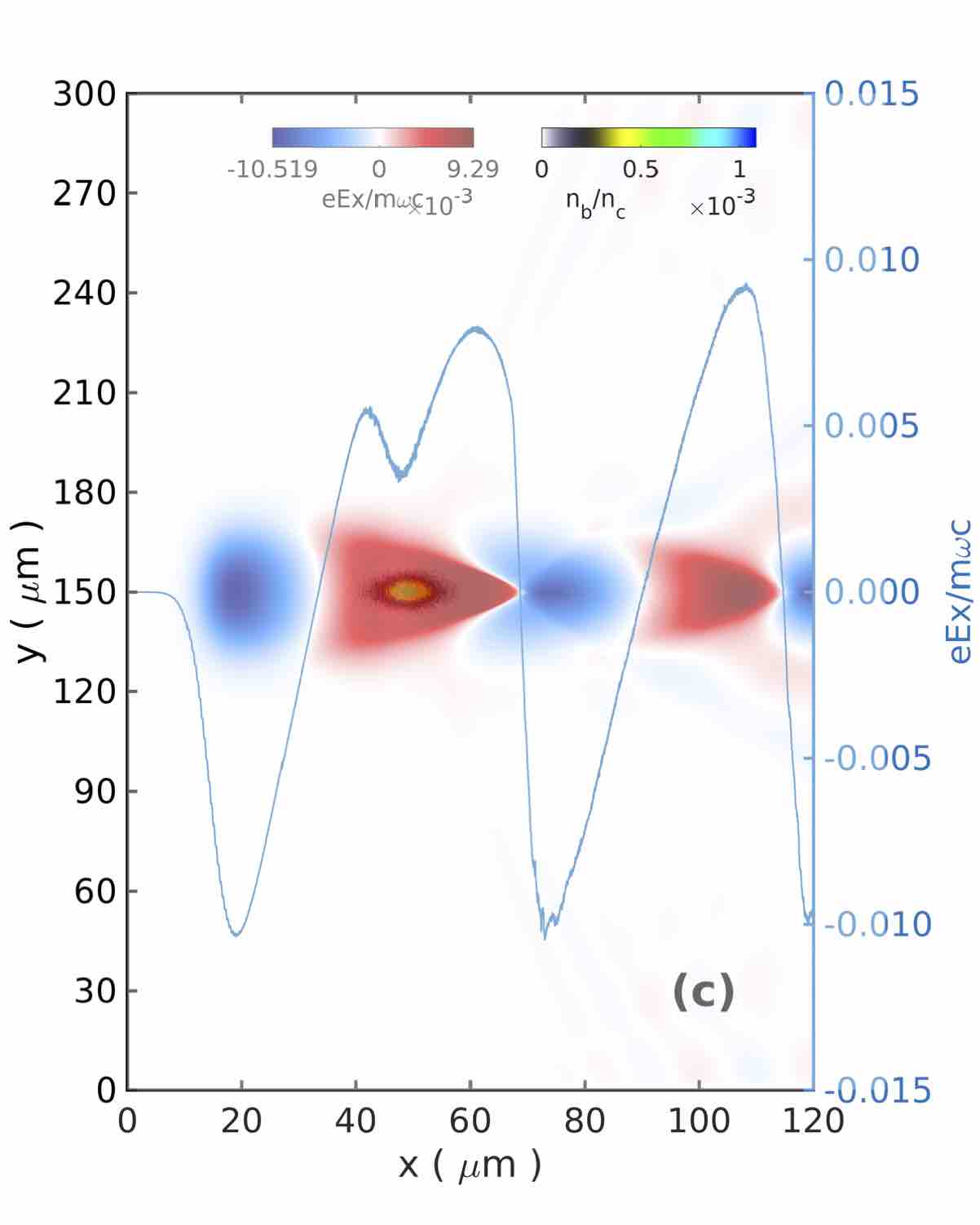}
\caption{}\label{fig:fig_c}
\end{subfigure}
\hskip 20pt
\begin{subfigure}[t]{.4\textwidth}
\centering
\vspace{-20pt}% set the real top as the top
\includegraphics[width=\linewidth]{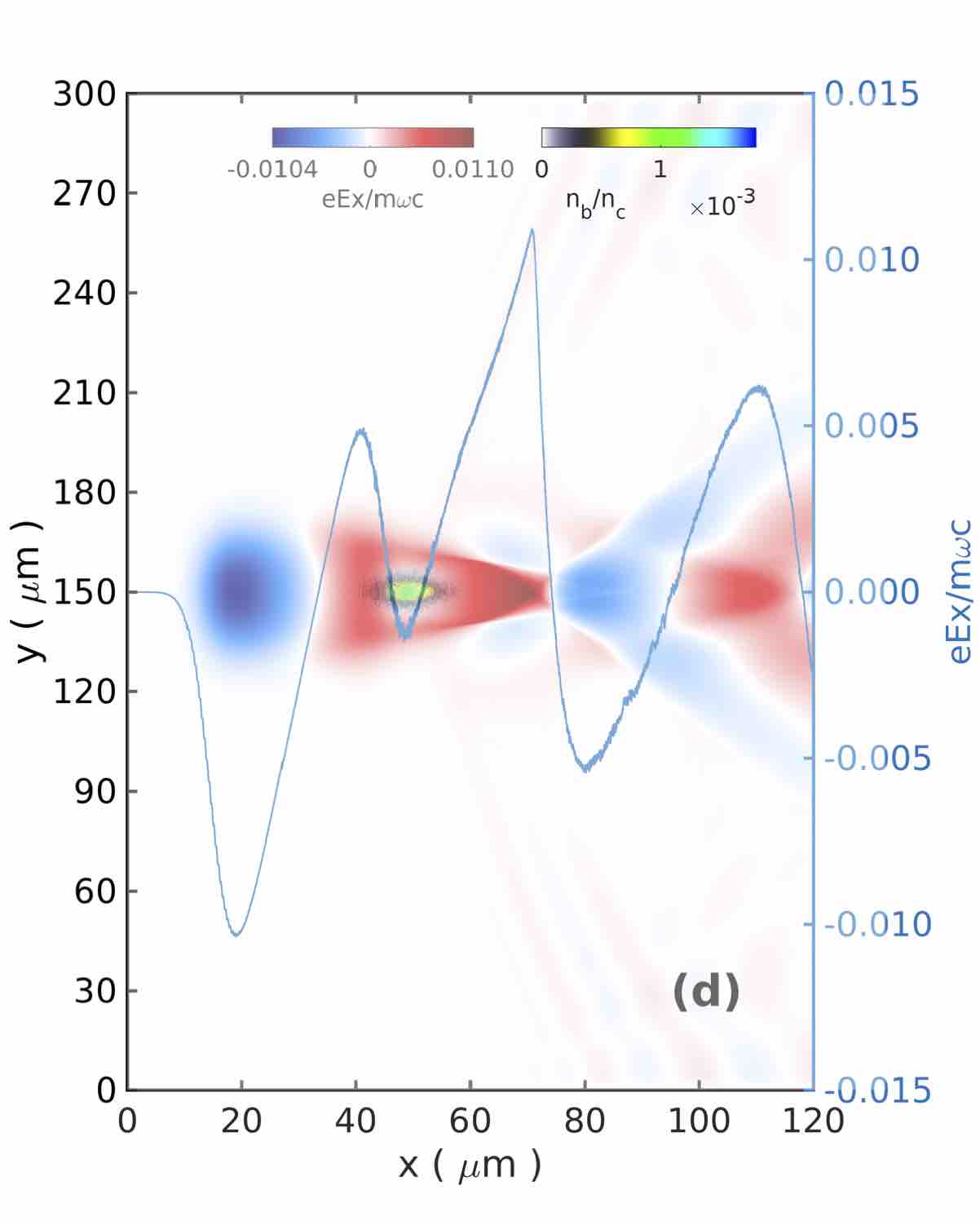}
%\caption{}\label{fig:fig_d}
\end{subfigure}
%

%\medskip

\begin{subfigure}[t]{.4\textwidth}
\centering
\vspace{-20pt}% set the real top as the top
\includegraphics[width=\linewidth]{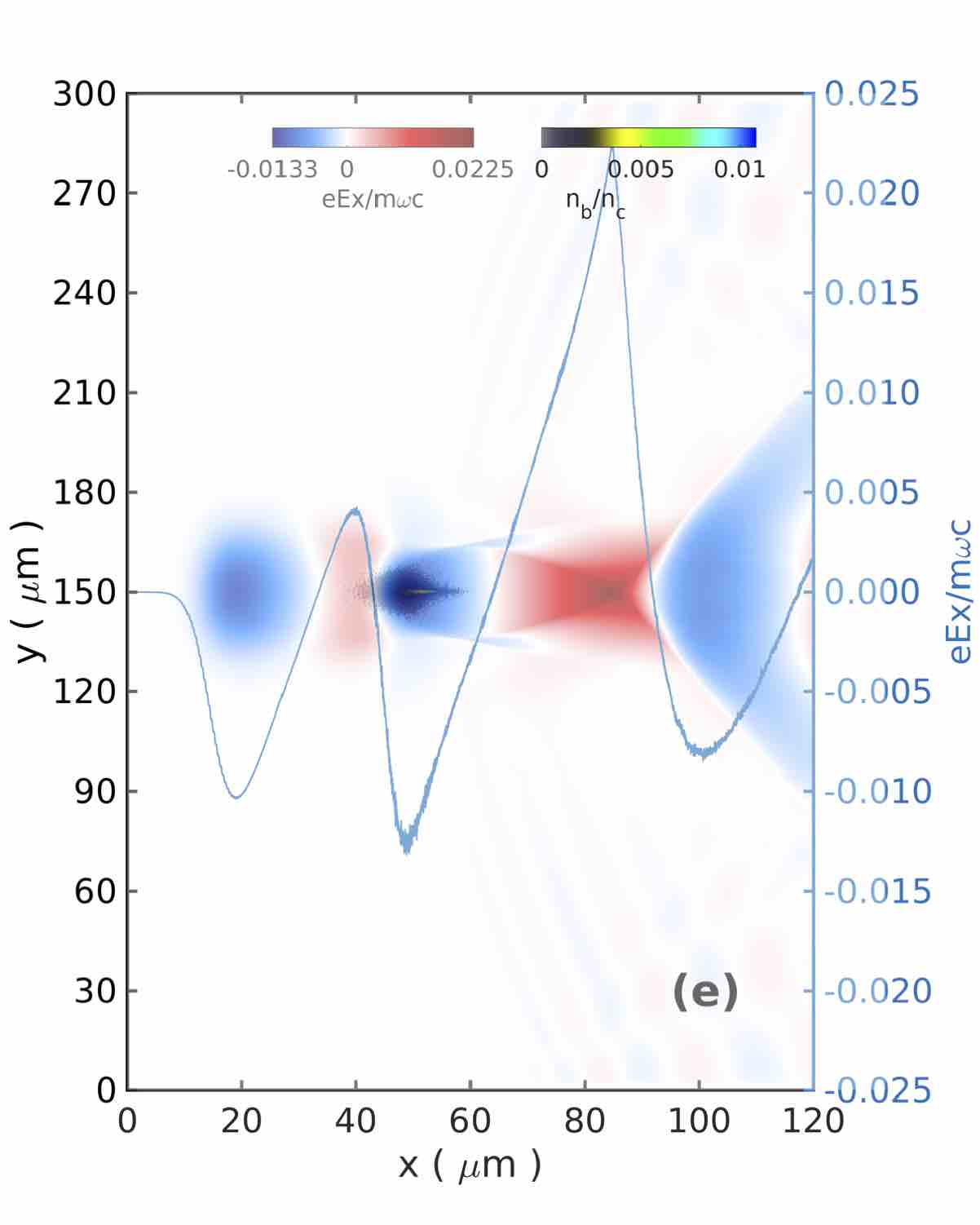}
%\caption{}\label{fig:fig_e}
\end{subfigure}
\hskip 20pt
\begin{minipage}[t]{.4\textwidth}
\vspace{-20pt}% set the real top as the top
\caption{\small Results of 3D PIC simulations displaying the effect of charge coupling on accelerating field in the booster stage. By increasing charge in the injected electron beam the accelerating field in the booster stage is deteriorated. The electron beam charge is (a) 10 pC, (b) 20 pC, (c) 30pC, (d) 50pC and (e) 100pC. The laser pulse is propagating from left to right hand side. The colorbar shows net accelerating field amplitude in normalized unit, and electron beam density normalized by critical density.}
\end{minipage}

\end{figure}

%=Figure:3===================================================

\begin{figure}
\centering
\begin{subfigure}[t]{.4\textwidth}
\centering
\includegraphics[width=\linewidth]{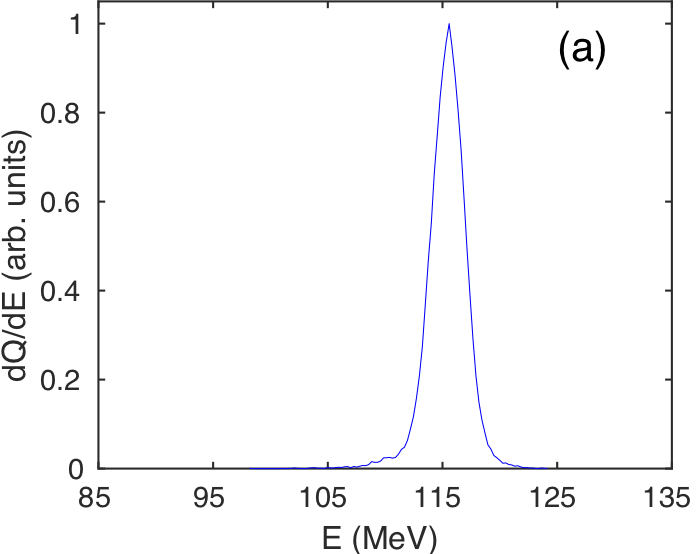}
%\caption{}\label{fig:fig_a}
\end{subfigure}
\hskip 20pt
\begin{subfigure}[t]{.4\textwidth}
\centering
\includegraphics[width=\linewidth]{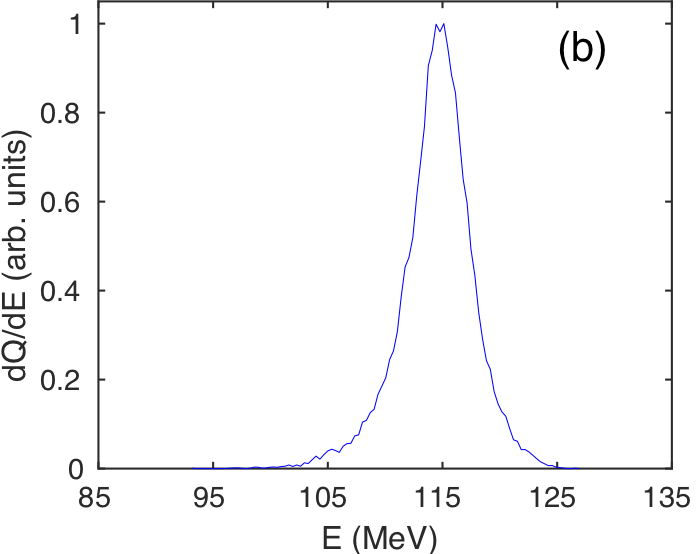}
%\caption{}\label{fig:fig_b}
\end{subfigure}

%\medskip

\begin{subfigure}[t]{.4\textwidth}
\centering
\vspace{10pt}% set the real top as the top
\includegraphics[width=\linewidth]{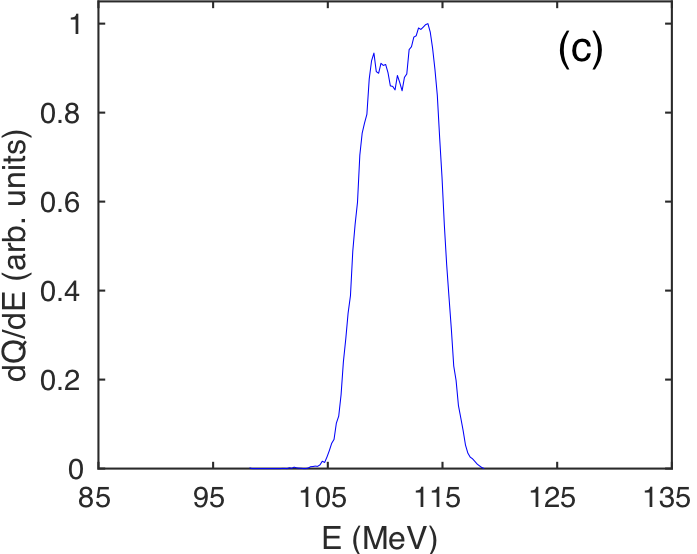}
%\caption{}\label{fig:fig_c}
\end{subfigure}
\hskip 20pt
\begin{subfigure}[t]{.4\textwidth}
\centering
\vspace{10pt}% set the real top as the top
\includegraphics[width=\linewidth]{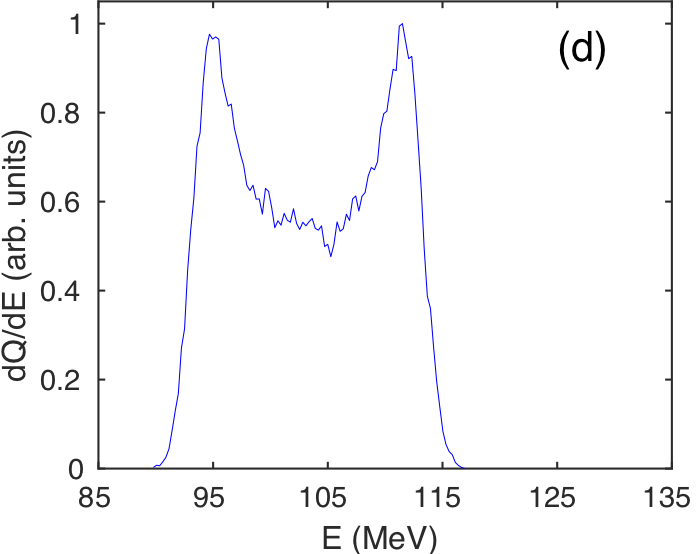}
%\caption{}\label{fig:fig_d}
\end{subfigure}
%

%\medskip

\begin{subfigure}[t]{.4\textwidth}
\centering
\vspace{25pt}% set the real top as the top
\includegraphics[width=\linewidth]{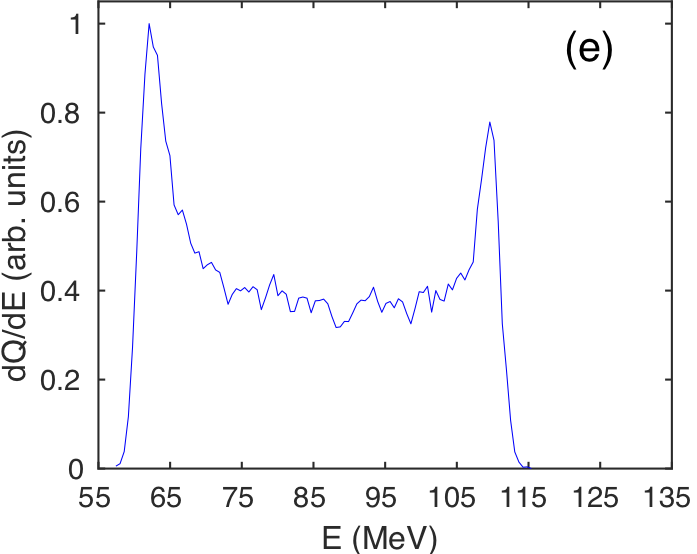}
%\caption{}\label{fig:fig_e}
\end{subfigure}
\hskip 20pt
\begin{minipage}[t]{.4\textwidth}
\vspace{15pt}% set the real top as the top
\caption{\small Results of 3D PIC simulations displaying the effect of charge coupling on the energy spectrum of the accelerated electron beam in the booster stage. The electron beam charge is (a) 10 pC, (b) 20 pC, (c) 30pC, (d) 50pC and (e) 100pC. Except electron beam charge all other parameter are exactly same in all the cases.}
\end{minipage}

\end{figure}

%=Figure:4===================================================

\begin{figure}
\centering
\begin{subfigure}[t]{.5\textwidth}
\vspace{-1.5cm}
\centering
\includegraphics[width=\linewidth]{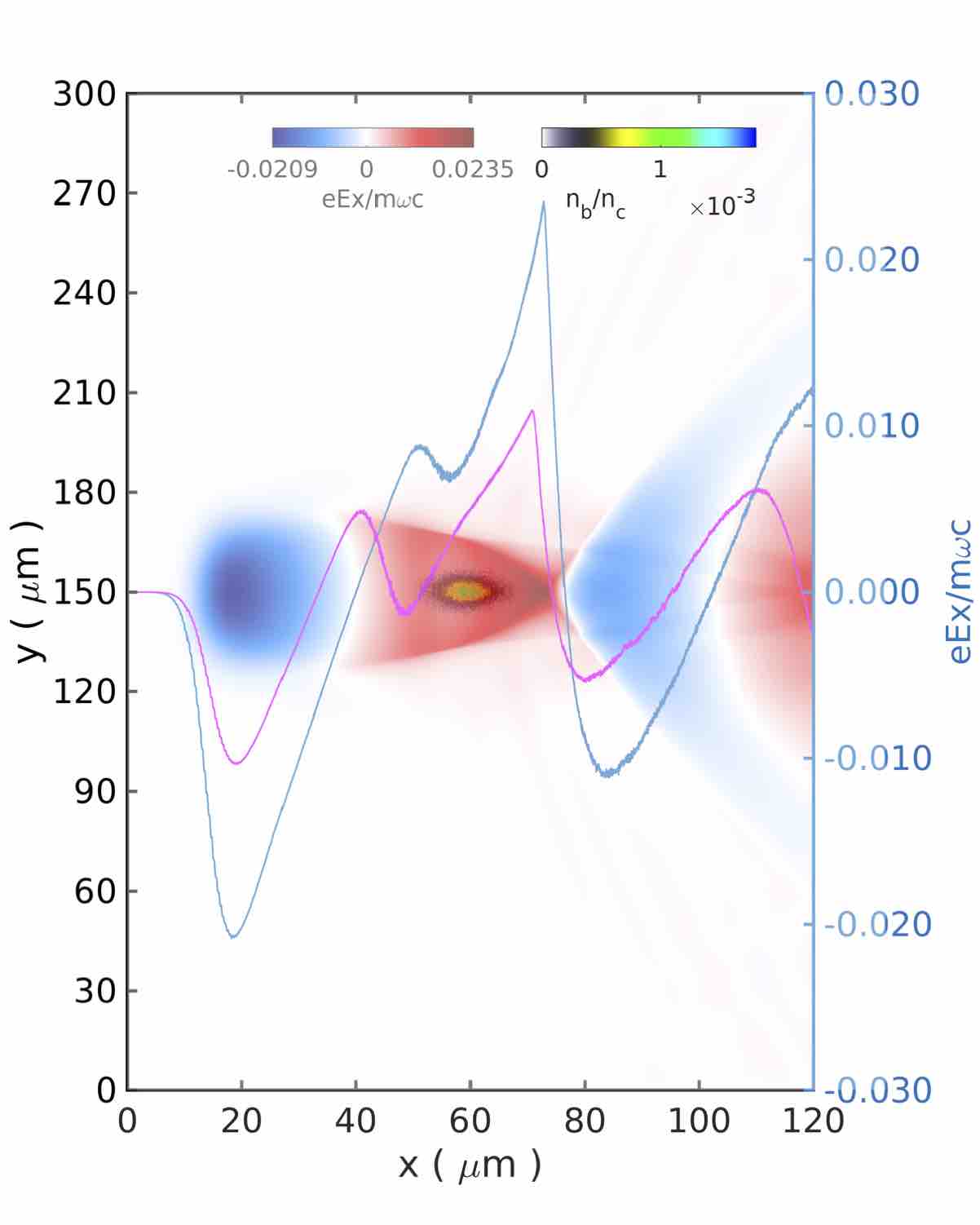}
\caption{}\label{fig:fig_a}
\end{subfigure}
%
%\hskip 20pt
\begin{subfigure}[t]{.5\textwidth}
\vspace{0.5cm}
\centering
\includegraphics[width=0.9\linewidth]{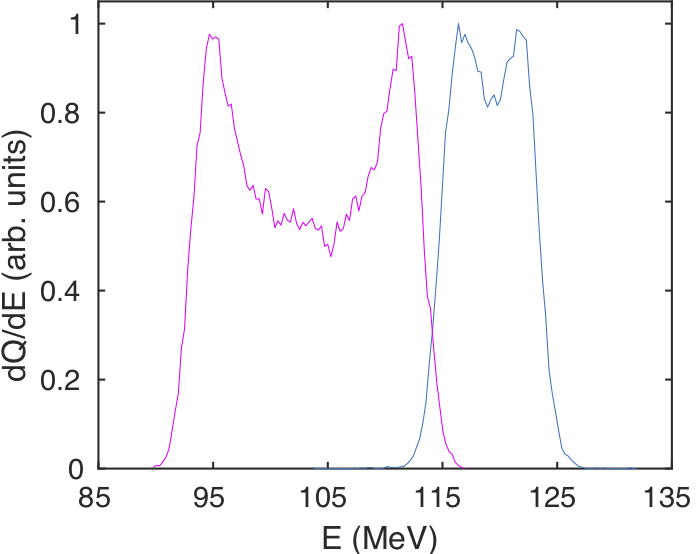}
\caption{}\label{fig:fig_b}
\end{subfigure}
\caption{\small Results of 3D PIC simulations displaying comparison of acceleration efficiency for $50 \ \rm pC$ electron bunch by increasing the laser pulse intensity in the booster stage. The laser pulse energy is increased from $4 \ \rm J$ to $10 \ \rm J$. All other parameters are the same as in the case of $4 \ \rm J$ laser pulse [Fig. 2(d)]. (a) The 2D pseudo colormap shows longitudinal field of the plasma wave (red-white-blue) superimposed by an accelerated electron beam (multi-color). The 1D line plot in blue color shows an on-axis effective accelerating field for $10 \ \rm J$ laser pulse, whereas, the 1D line plot in magenta color shows an effective accelerating field for $4 \ \rm J$ laser pulse. (b) Blue color display the energy spectrum of an externally injected electron beam with $10 \ \rm J$ laser pulse in the booster stage, whereas, magenta color display energy spectrum with $4 \ \rm J$ laser pulse. In (a) laser pulse is propagating from right to left hand side. The colorbar shows net accelerating field amplitude in normalized unit, and electron beam density normalized by critical density.}
\end{figure}

%=Figure:5===================================================

\begin{figure}
\centering
\begin{subfigure}[t]{.5\textwidth}
\vspace{-1.5cm}
\centering
\includegraphics[width=\linewidth]{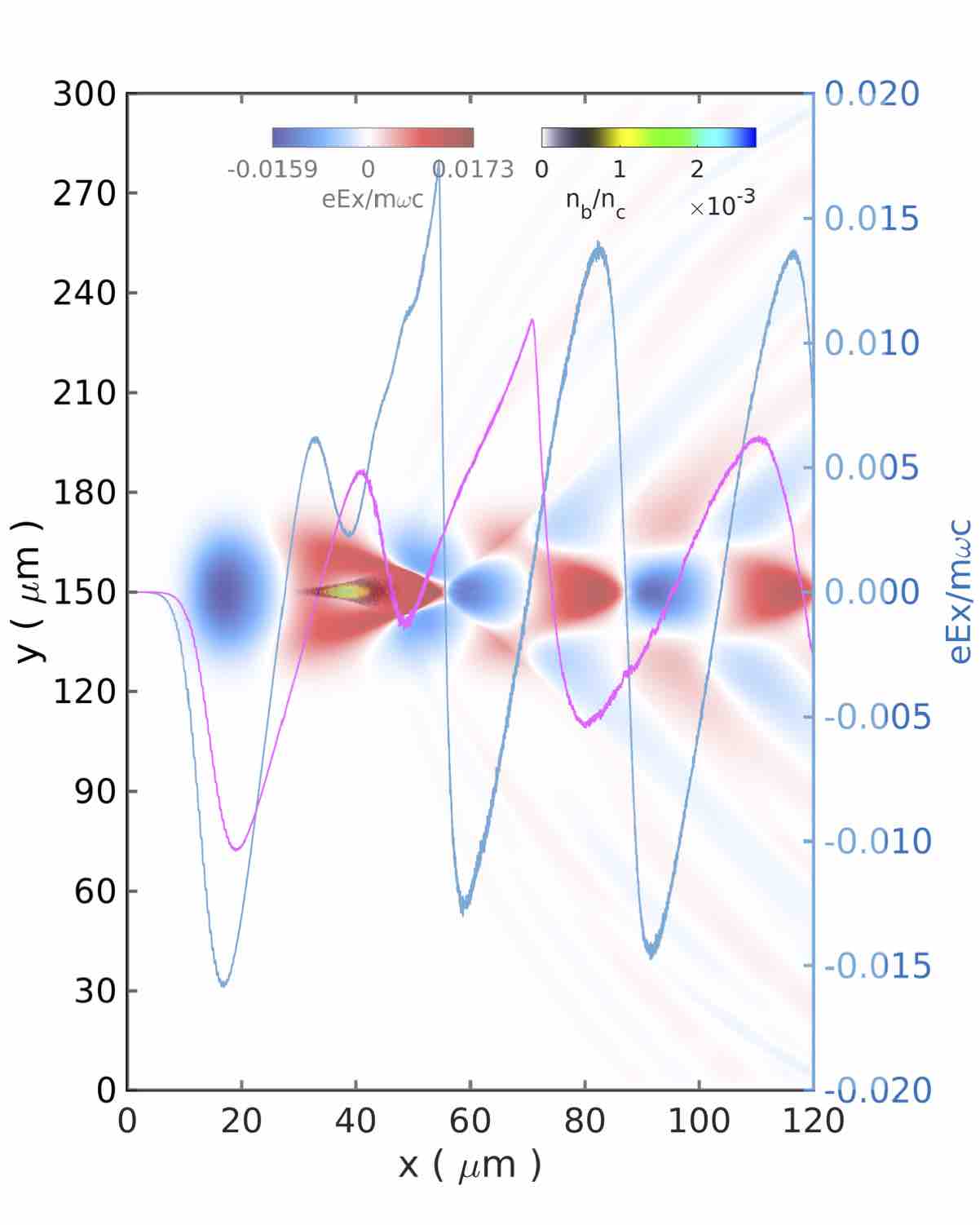}
\caption{}\label{fig:fig_a}
\end{subfigure}
%
%\hskip 20pt
\begin{subfigure}[t]{.5\textwidth}
\vspace{0.5cm}
\centering
\includegraphics[width=0.9\linewidth]{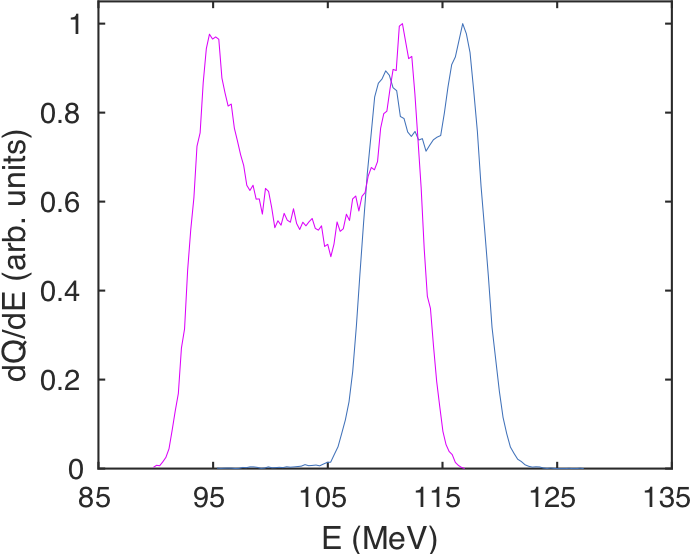}
\caption{}\label{fig:fig_b}
\end{subfigure}
\caption{\small Results of 3D PIC simulations displaying comparison of acceleration efficiency for $50 \ \rm pC$ electron bunch by increasing the plasma density in the booster stage. The plasma density is increased from $5 \times 10^{17}cm^{-3}$ to $1 \times 10^{18}cm^{-3}$ . All other parameters are the same as in the case of $5 \times 10^{17}cm^{-3}$ [Fig. 2(d)]. (a) The 2D pseudo colormap shows longitudinal field of the plasma wave (red-white-blue) superimposed by an accelerated electron beam (multi-color). The 1D line plot in blue color shows an on-axis effective accelerating field for $1 \times 10^{18}cm^{-3}$, whereas, the 1D line plot in magenta color shows an effective accelerating field for $5 \times 10^{17}cm^{-3}$. (b) Blue color display the energy spectrum an externally injected electron beam with plasma density of $1 \times 10^{18}cm^{-3}$ in the booster stage, whereas, magenta color display energy spectrum with plasma density of $5 \times 10^{17}cm^{-3}$. In (a) laser pulse is propagating from right to left hand side. The colorbar shows net accelerating field amplitude in normalized unit, and electron beam density normalized by critical density.}
\end{figure}

%=Figure:6===================================================

\begin{figure}
\centering
\begin{subfigure}[t]{.5\textwidth}
\vspace{-1.5cm}
\centering
\includegraphics[width=\linewidth]{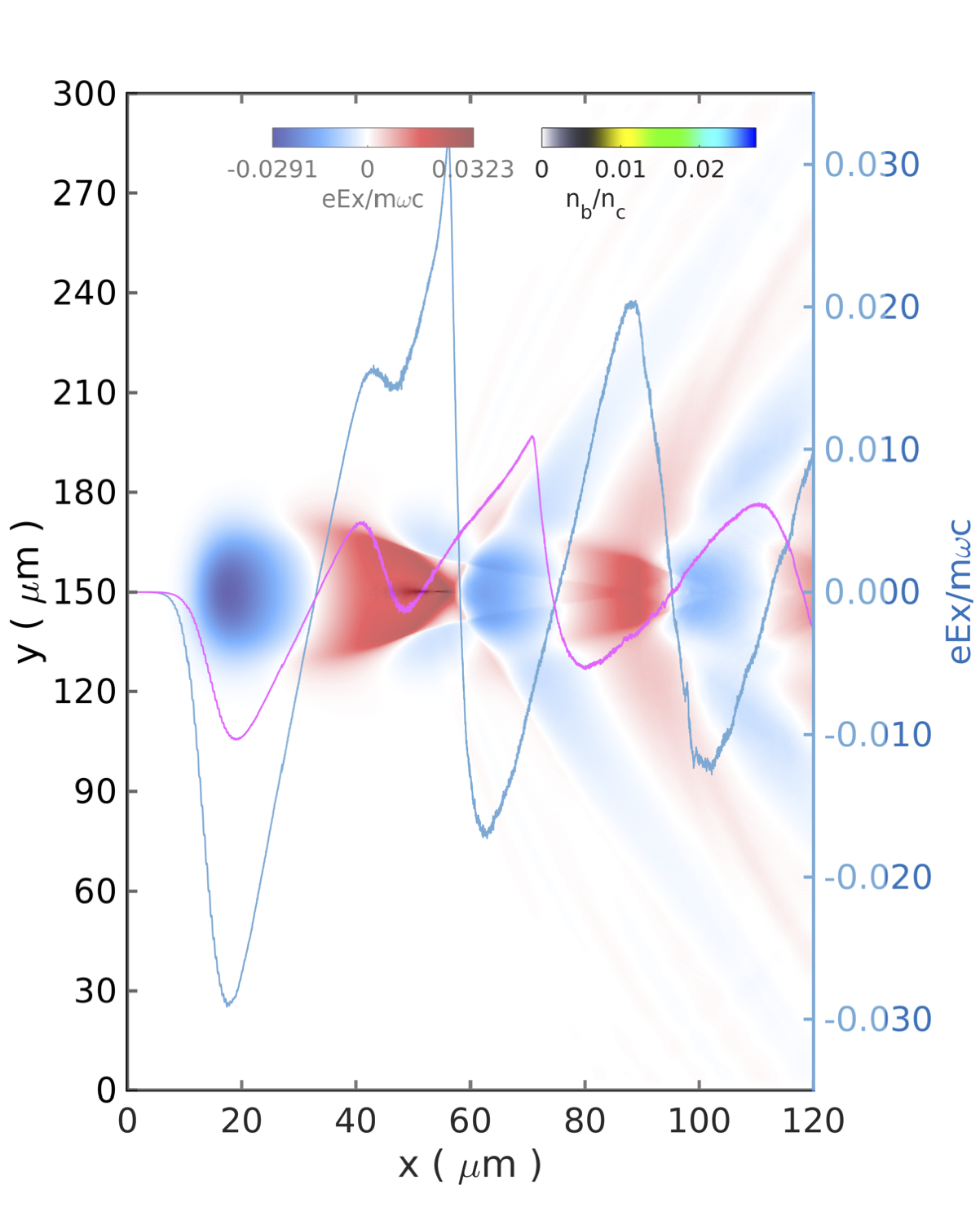}
\caption{}\label{fig:fig_a}
\end{subfigure}
%
%\hskip 20pt
\begin{subfigure}[t]{.5\textwidth}
\vspace{0.5cm}
\centering
\includegraphics[width=0.9\linewidth]{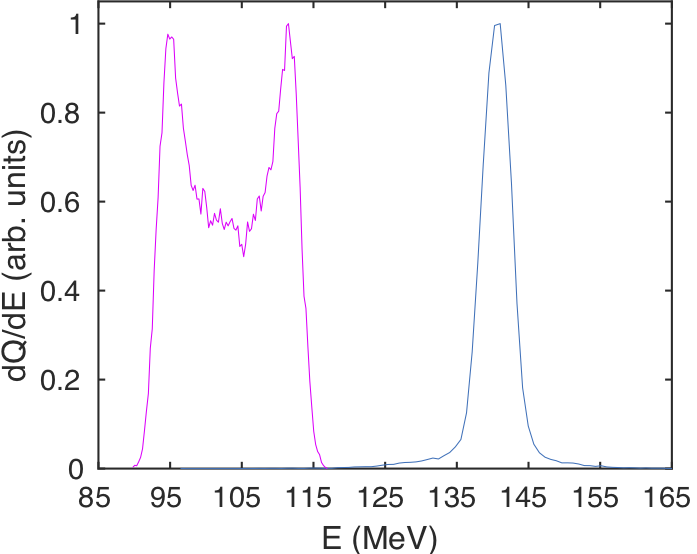}
\caption{}\label{fig:fig_b}
\end{subfigure}
\caption{\small Results of 3D PIC simulations displaying comparison of acceleration efficiency for $50 \ \rm pC$ electron bunch by increasing the laser pulse intensity as well as plasma density in the booster stage. All other parameters are the same as in the case of Fig. 2(d). (a) The 2D pseudo colormap shows longitudinal field of the plasma wave (red-white-blue) superimposed by an accelerated electron beam (multi-color). The 1D line plot in blue color shows an on-axis effective accelerating field for high intensity and plasma density, whereas, the 1D line plot in magenta color shows an effective accelerating field for low intensity and plasma density. (b) Blue color display the energy spectrum an externally injected electron beam for high intensity and plasma density, whereas, magenta color display energy spectrum an externally injected electron beam for low intensity and plasma density. In (a) laser pulse is propagating from right to left hand side. The colorbar shows net accelerating field amplitude in normalized unit, and electron beam density normalized by critical density.}
\end{figure}

%=Figure:7===================================================

\begin{figure}
\centering
\begin{subfigure}[t]{.5\textwidth}
\vspace{-1.5cm}
\centering
\includegraphics[width=\linewidth]{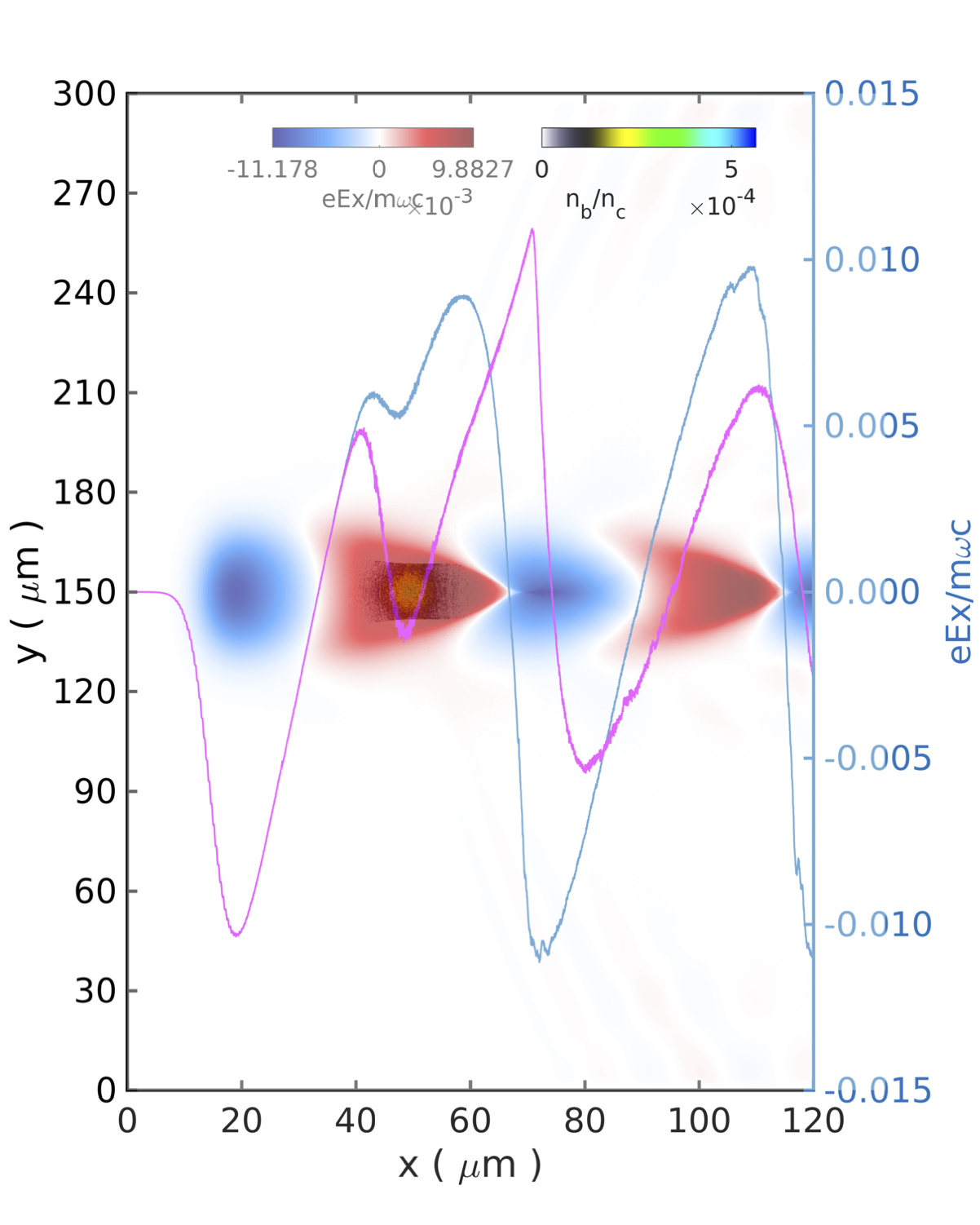}
\caption{}\label{fig:fig_a}
\end{subfigure}
%
%\hskip 20pt
\begin{subfigure}[t]{.5\textwidth}
\vspace{0.5cm}
\centering
\includegraphics[width=0.9\linewidth]{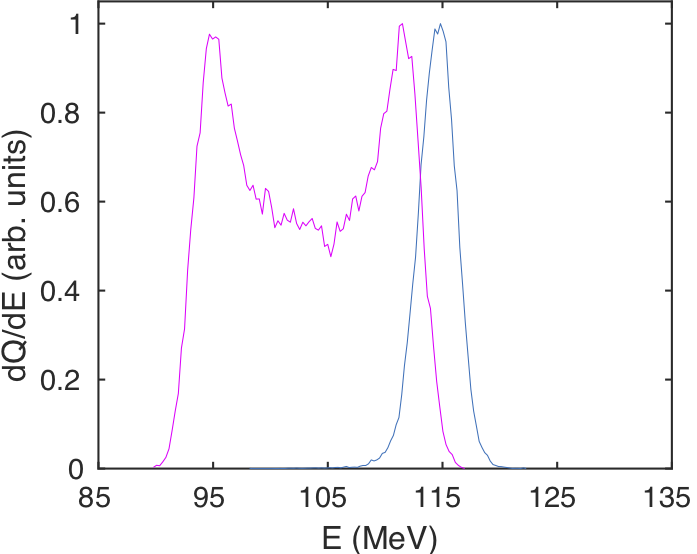}
\caption{}\label{fig:fig_b}
\end{subfigure}
\caption{\small Results of 3D PIC simulations displaying comparison of acceleration efficiency for $50 \ \rm pC$ electron bunch by increasing the transverse size of the electron beam. The radius $(r_{b})$ of the electron beam is increased from $5 \ \mu m$ to $10 \ \mu m$. All other parameters are the same as in the case of $5 \ \mu m$ electron bunch [Fig. 2(d)]. (a) The 2D pseudo colormap shows longitudinal field of the plasma wave (red-white-blue) superimposed by an accelerated electron beam (multi-color). The 1D line plot in blue color shows an on-axis effective accelerating field for $r_{b}=10 \ \mu m$, whereas, the 1D line plot in magenta color shows an effective accelerating field for $r_{b}=5 \ \mu m$. (b) Blue color display the energy spectrum an externally injected electron beam of radius $r_{b}=10 \ \mu m$, whereas, magenta color display energy spectrum an externally injected electron beam of radius $r_{b}=5 \ \mu m$. In (a) laser pulse is propagating from right to left hand side. The colorbar shows  net accelerating field amplitude in normalized unit, and electron beam density normalized by critical density.}
\end{figure}

%=Figure:8===================================================

\begin{figure}
\centering
\includegraphics[width=1.1\linewidth]{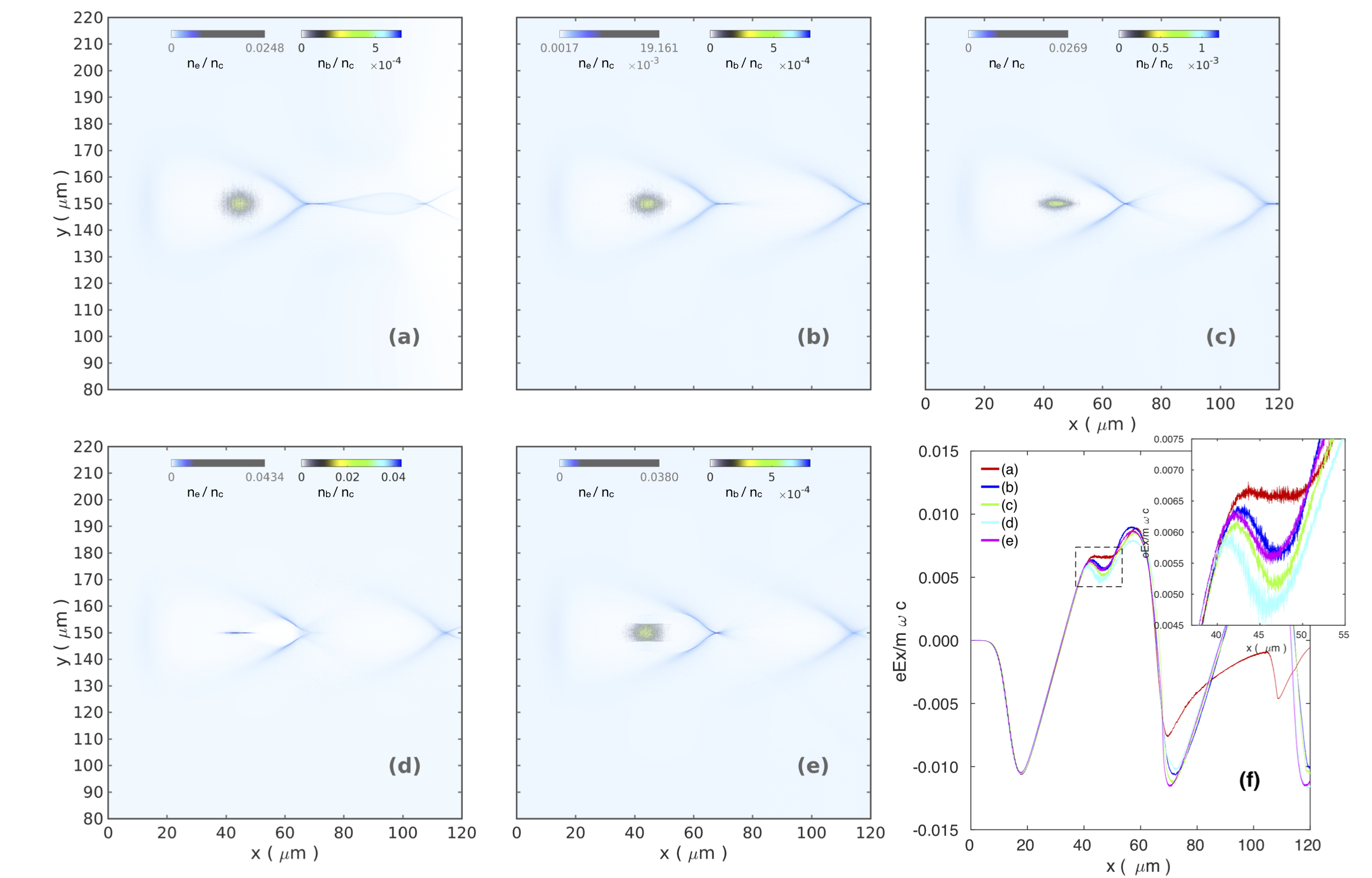}
\caption{\small Result of 3D PIC simulation illustrating the evolution of an externally injected electron beam in the booster stage plasma wave. (a-e) show dynamic focusing and defocusing of the electron beam inside the plasma wave, and (f) shows effect of the beam dynamics on the net accelerating field of the plasma wave. In (f) each line plot of the net accelerating field is corresponding to the electron beam dynamic evoluiton in (a-e). Inset shows the magnified view of the net accelerating field under the dotted rectangle. Time corresponding to each snap shot is: (a) 0.4 ps, (b) 0.6 ps, (c) 0.8 ps, (d) 1.0 ps and (e) 1.5 ps. The colorbar shows plasma density normalized by critical density, and electron beam density normalized by critical density.}\label{fig:8}
\end{figure}

%=Figure:9===================================================

\begin{figure}
\centering
\begin{subfigure}[t]{.4\textwidth}
\centering
\includegraphics[width=\linewidth]{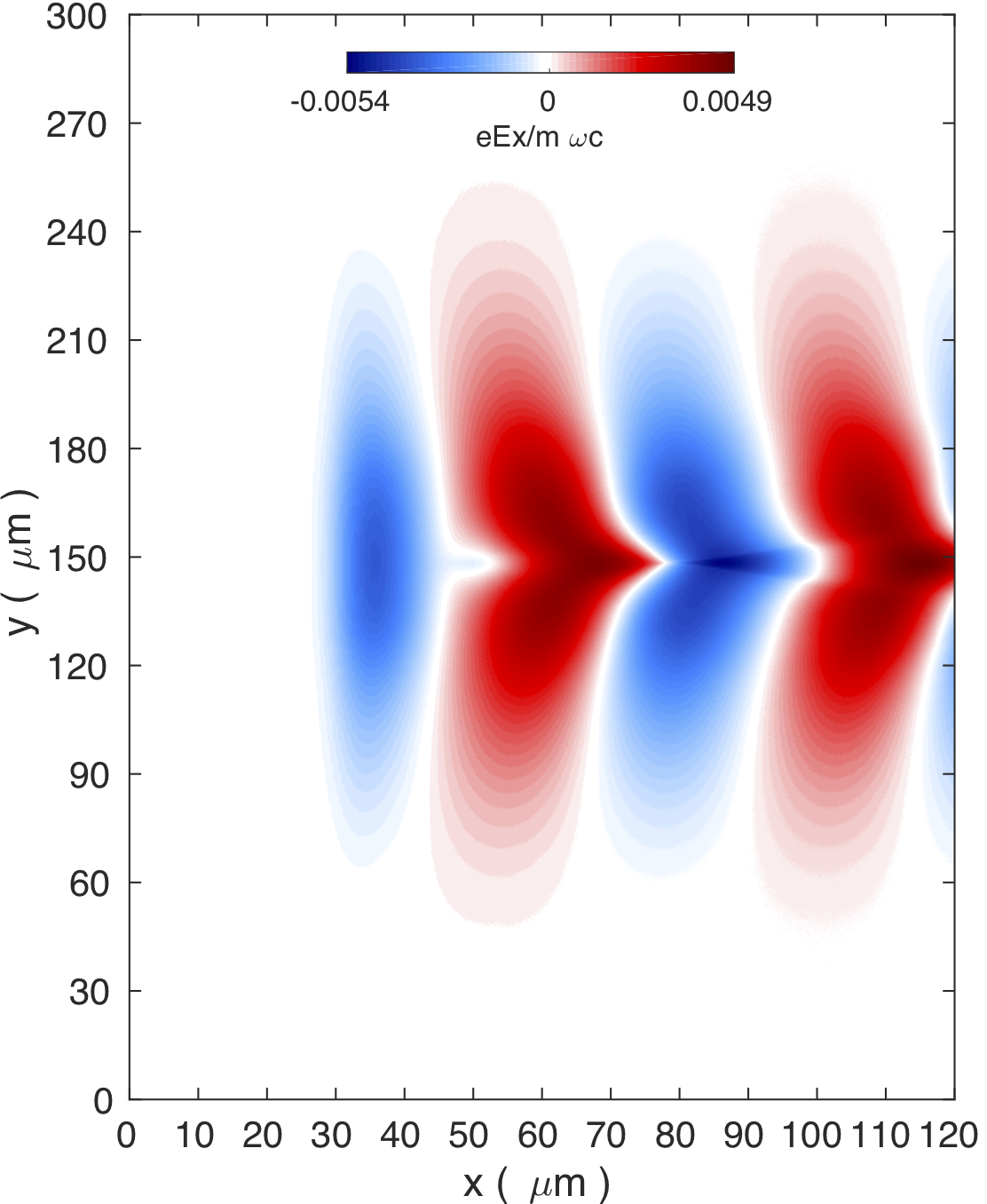}
\caption{}\label{fig:fig_a}
\end{subfigure}
\hskip 10pt
\begin{subfigure}[t]{.4\textwidth}
\centering
\includegraphics[width=\linewidth]{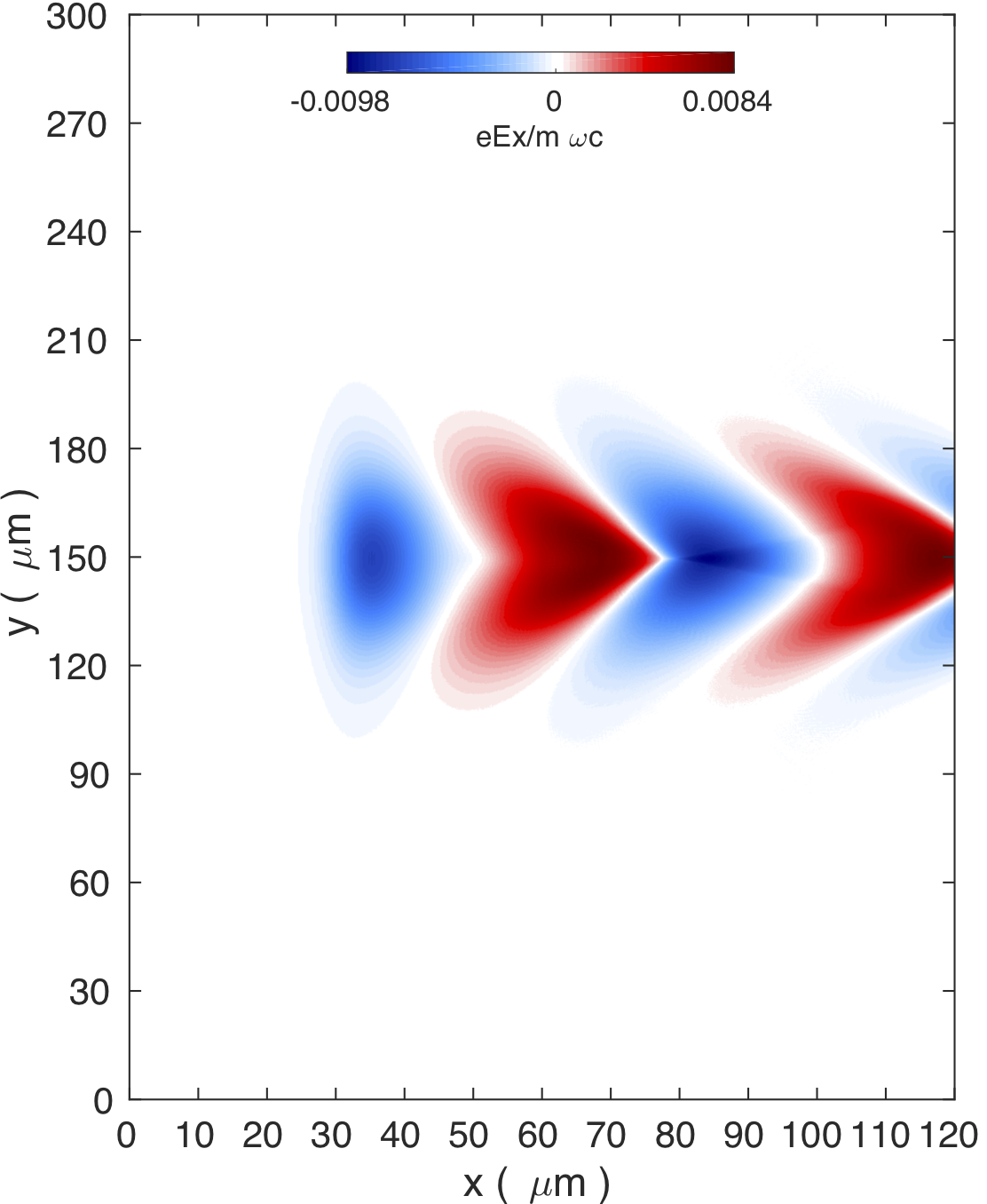}
\caption{}\label{fig:fig_b}
\end{subfigure}
\caption{\small Results of 2D PIC simulations displaying the evolution of the accelerating field of the plasma wave in the booster stage. (a) Corresponds to the propagation of the laser pulse in uniform plasma density of $5 \times 10^{17}cm^{-3}$, whereas, (b) corresponds to the propagation of the laser pulse in parabolic plasma channel. The plasma density along the axis is  $5 \times 10^{17}cm^{-3}$ and depth $0.5$ i.e. maximum density of $1 \times 10^{18}cm^{-3}$ on the periphery of the channel. Except plasma density profile all other parameters are exactly same in both cases. Laser pulse is propagating from right to left hand side. The accelerating field is shown at $30 \ ps$ or $ 9 \ mm$ propagation length. The colorbar shows net accelerating field amplitude in normalized unit.}

\end{figure}

%=Figure:10===================================================

\begin{figure}
\centering
\includegraphics[width=0.55\linewidth]{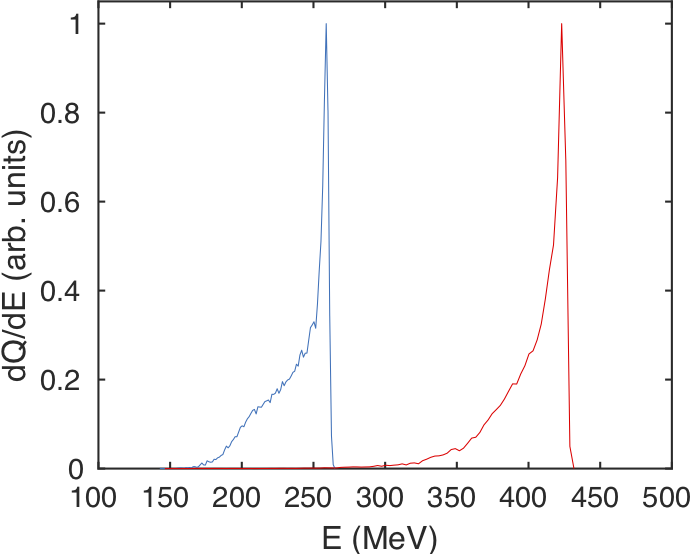}
\caption{\small Results of 2D PIC simulations displaying the energy spectrum of an externally injected electron beam ($100 \ MeV$) co-propagating with the plasma wave in the booster stage. The blue color graph shows energy gain by the electron beam in the booster stage consist of uniform plasma density, whereas, the red color graph shows energy spectrum in the booster stage consist of parabolic plasma channel. Efficient acceleration and energy gain in uniform density plasma is limited due to refractive loses of the laser pulse in the transverse direction. The energy spectrum is shown at $30 \  ps$ or $9 \ mm$ propagation length.}
\end{figure}

%===========================================================================


\begin{thebibliography}{99}
\bibitem{comment}
Here, the word efficiency define the acceleration ability of the plasma wave without being effected by the electron beam. It should not be confused with the energy transfer ratio between the plasma wave and the electron beam.
\end{thebibliography}
\end{document}